%% file: main.tex
\documentclass[10pt, conference, letterpaper]{IEEEtran}
\IEEEoverridecommandlockouts
\usepackage{cite}
\usepackage{amsmath,amssymb,amsfonts}
\usepackage{graphicx}
\usepackage{textcomp}
\usepackage{xcolor}
\def\BibTeX{{\rm B\kern-.05em{\sc i\kern-.025em b}\kern-.08em
    T\kern-.1667em\lower.7ex\hbox{E}\kern-.125emX}}
    
\usepackage{amsthm,amssymb,graphicx,multirow,amsmath,color,amsfonts}
\usepackage[update,prepend]{epstopdf}
\usepackage[latin1]{inputenc}
\usepackage{tikz}
\usepackage{bbm} 
\usepackage{pdfpages}
\usepackage{tabulary}
\usepackage{multirow}
\usepackage{comment}
\usepackage{graphicx}
\usepackage{amsthm}
\usepackage{caption}
\usepackage{subcaption}
\usepackage{mathtools}
\usepackage{epstopdf}
\usepackage{comment}
\usepackage{adjustbox}
\usepackage{tcolorbox}
\usepackage{subcaption}
\usepackage{caption}
\usepackage{algorithm}
\usepackage{algpseudocode}
\usepackage[noframe]{showframe} % remove this to show the frame and see where you are violating this
\usepackage{array,ragged2e}
\newcolumntype{P}[1]{>{\RaggedRight\arraybackslash}p{#1}}

\newcommand{\subparagraph}{}
\long\def\/*#1*/{}

\include{notation}
\allowdisplaybreaks
\begin{document}

%\titleformat{\subsubsection}
%  {\normalfont\normalsize\textbf}{\thesubsubsection.}{1em}{}

\graphicspath{{./Figures/}}

%\title{xApp Repository Function: A Scalable Authentication, Authorization and Discovery Framework for OpenRAN}
%\title{Securing 5G OpenRAN with the xApp Repository Function: A Scalable Authentication, Authorization and Discovery Framework}
\title{Securing 5G OpenRAN with a Scalable Authorization Framework for xApps}
%\title{xApp Repository Function: A Novel Security Framework for 5G OpenRAN}
%\title{Securing 5G with the xApp Repository Function: A Scalable Authentication, Authorization and Discovery Framework for OpenRAN}

\author{\IEEEauthorblockN{Tolga O. Atalay\IEEEauthorrefmark{1}, Sudip Maitra\IEEEauthorrefmark{1}, Dragoslav Stojadinovic\IEEEauthorrefmark{2}, Angelos Stavrou\IEEEauthorrefmark{1}\IEEEauthorrefmark{2}, Haining Wang\IEEEauthorrefmark{1}}
\IEEEauthorblockA{\IEEEauthorrefmark{1}Department of Electrical and Computer Engineering, Virginia Tech, USA\\
\IEEEauthorrefmark{2}Kryptowire, LLC, McLean, VA, USA\\
%Email: \{tolgaoa, stojadinovic, afamili, angelos, hnw\}@vt.edu
Email: tolgaoa@vt.edu,
smaitra@vt.edu,
dstojadinovic@kryptowire.com,
angelos@vt.edu,
hnw@vt.edu
}

\thanks{This material is based on research sponsored by Defense Advanced Research Projects Agency (DARPA) under agreement number HR001120C0155. The views, opinions, and/or findings contained in this article are those of the author(s) and should not be interpreted as representing the official views or policies, either expressed or implied, of the Defense Advanced Research Projects Agency or the Department of Defense.}
}

\newcommand\acceptedpaper{%
  \footnotsize This paper has been accepted to IEEE International Conference on Computer Communications (INFOCOM) 2023.
  }

  \newcommand\acceptednotice{%
\begin{tikzpicture}[remember picture,overlay]
\node[anchor=south,yshift=10pt] at (current page.north) {\fbox{\parbox{\dimexpr\textwidth-\fboxsep-\fboxrule\relax}{\copyrighttext}}};
\end{tikzpicture}%
}

\newcommand\copyrighttext{%
  \footnotesize \textcopyright 2023 IEEE. Personal use of this material is permitted.
  Permission from IEEE must be obtained for all other uses, in any current or future
  media, including reprinting/republishing this material for advertising or promotional
  purposes, creating new collective works, for resale or redistribution to servers or
  lists, or reuse of any copyrighted component of this work in other works.
  %DOI: \href{<http://tex.stackexchange.com>}{<DOI No.>}
  }
\newcommand\copyrightnotice{%
\begin{tikzpicture}[remember picture,overlay]
\node[anchor=south,yshift=10pt] at (current page.south) {\fbox{\parbox{\dimexpr\textwidth-\fboxsep-\fboxrule\relax}{\copyrighttext}}};
\end{tikzpicture}%
}

\maketitle

%\acceptedpaper
\copyrightnotice

%\thispagestyle{empty}
%\pagestyle{empty}

%9 lines to not exceed limit

\begin{abstract}
The ongoing transformation of mobile networks from proprietary physical network boxes to virtualized functions and deployment models has led to more scalable and flexible network architectures capable of adapting to specific use cases. As an enabler of this movement, the OpenRAN initiative promotes standardization allowing for a vendor-neutral radio access network with open APIs. Moreover, the O-RAN Alliance has begun specification efforts conforming to OpenRAN's definitions. This includes the near-real-time RAN Intelligent Controller (RIC) overseeing a group of extensible applications (xApps). The use of these potentially untrusted third-party applications introduces a new attack surface to the mobile network plane with fundamental security and system design requirements that are yet to be addressed. To secure the 5G O-RAN xApp model, we introduce the xApp Repository Function (XRF) framework, which implements scalable authentication, authorization, and discovery for xApps. We first present the framework's system design and implementation details, followed by operational benchmarks in a production-grade containerized environment. The evaluation results, centered on active processing and operation times, show that our proposed framework can scale efficiently in a multi-threaded Kubernetes microservice environment and support a large number of clients with minimal overhead.
\end{abstract}
%the abstract is already uploaded

%%%%%%%%%%%%%%%%%%%%%%%%%%%%%%%%%
\section{Introduction} \label{sec:intro}
\input{introduction/texfile_v2.tex}

%%%%%%%%%%%%%%%%%%%%%%%%%%%%%%%%%
\section{Related Work} \label{sec:relwork}
\input{relworks/texfile_v2.tex}
%%%%%%%%%%%%%%%%%%%%%%%%%%%%%%%%%
%\section{Background Information} \label{sec:background}
%\input{background/texfile.tex}
%%%%%%%%%%%%%%%%%%%%%%%%%%%%%%%%%
\section{Threat Model} \label{sec:thrmdl}
\input{thrmdl/texfile}

%\input{thrmdl/texfile_sudip}
%%%%%%%%%%%%%%%%%%%%%%%%%%%%%%%%%
\section{XRF Framework} \label{sec:frmwdsgn}
\input{frmwdsgn/texfile_v2}

%%%%%%%%%%%%%%%%%%%%%%%%%%%%%%%%%
%\section{Experimental Setup} \label{sec:expsetup}
%\input{expsetup/texfile}
%%%%%%%%%%%%%%%%%%%%%%%%%%%%%%%%%
\section{Performance Evaluation} \label{sec:perfeval}
\input{perfeval/texfile}
%%%%%%%%%%%%%%%%%%%%%%%%%%%%%%%%%
\section{Conclusion} \label{sec:conc}
\input{conc/texfile}
%%%%%%%%%%%%%%%%%%%%%%%%%%%%%%%%%

\newpage
\bibliographystyle{IEEEtran}
\bibliography{references}

\end{document}

%% file: introduction/texfile_v2.tex
%These 5G networks will be servicing a wide range of use cases with versatile demands and specific quality of service (QoS) requirements. 

%To sustain such a system, offerings will be carried out over logically isolated set of radio access and compute resources denoted as network slices within the 5G ecosystem. Each network slice will be specifically tailored to support different use cases such as enhanced mobile broadband (eMBB), ultra reliable low latency communication (URLLC), massive internet of things (mIoT) and a larger selection of customized slice service types (SST)~\cite{3gpp23501} created by mobile virtual network operators (MVNOs). 

The deployment of the next generation of mobile networks is underway. Mobile network functions which existed as physical network functions (PNFs) on proprietary hardware, are now implemented as virtual network functions (VNFs) on commercial off-the-shelf (COTS) servers. This transition enables the scalable and flexible deployment of the 5G radio access networks (RAN) and  5G core (5GC) in cloud-supported virtualized environments. To that end, the OpenRAN initiative, with the O-RAN Alliance~\cite{ORANALLI41online,9579445} as a leader, proposes the disaggregation and softwarization of the RAN by employing a functionality split where virtualized logical entities host different layers of the radio stack~\cite{3gpp38801}. Additionally, by steering the 5G deployment effort towards the adoption of open RAN interfaces with standardized APIs, it promotes a vendor-neutral integration of the 5G RAN.

% that will allow improvement efforts from multiple communities to seamlessly come together. %This collaboration will accelerate the delivery of the infrastructure necessary to support the versatility that 5G networks need.

\begin{figure*}[h!]
    \centering
    \includegraphics[width=0.95\textwidth,trim={1cm 8cm 3.cm 5.3cm},clip]{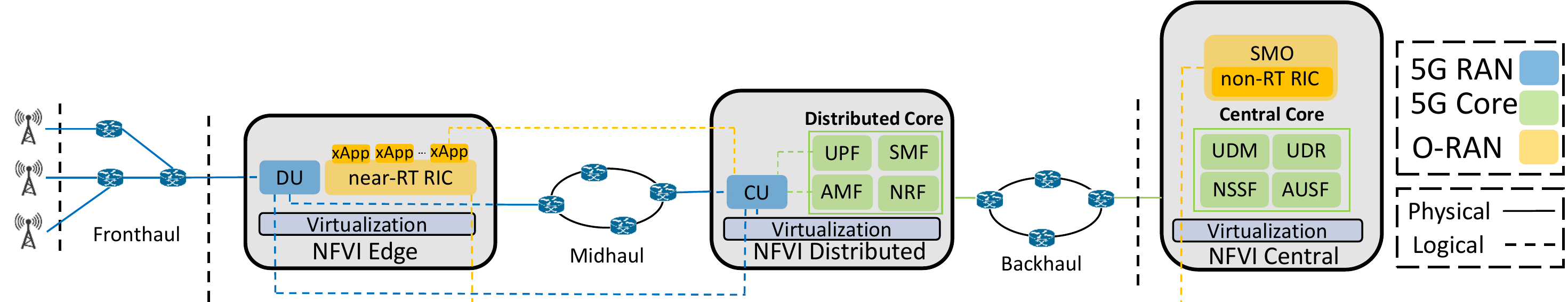}
     \caption{High level overview of the 5G ecosystem with O-RAN augmentation}
     \label{fig:highlvl5g}
\end{figure*}

Furthermore, the O-RAN architecture introduces a new software-defined networking (SDN) controller called the near real-time RAN Intelligent Controller (near-RT RIC), which hosts various RAN-related utility VNFs in the form of extensible applications (xApps). Through xApps, mobile virtual network operators (MVNOs) are offered a high degree of freedom in accessing and manipulating information in the radio stack in a software-defined manner. Every gNB within the O-RAN framework is fitted with an agent~\cite{orane2gen}, allowing it to interact with the near-RT RIC using a custom access protocol~\cite{orane2ap}. Fig.~\ref{fig:highlvl5g} shows the high-level integration of O-RAN into the 5G architecture.
The scenario depicts a RAN functionality split where the lower layers are implemented in the distributed unit (DU) while the upper layers are inside the central unit (CU). %Both the DU and CU are softwarized entities deployed at micro-datacentres, made up of COTS hardware. 
The O-RAN near-real time RIC is deployed at the same hierarchy as the DU, where xApps will subscribe to information from the radio stack. The management back-end of O-RAN is deployed in the central cloud along with the 5G core components. %Certain 5G core components have been deployed in a more distributed fashion while others have been centralized depending on operational requirements.

%The scenario depicts the 7.2 functionality split~\cite{3gpp38806} adopted by O-RAN in the 5G RAN protocol stack. As a result of this, the first three layers are implemented closer to the fronthaul in the distributed unit (DU) while the upper two layers are closer to the backhaul inside the central unit (CU). Both the DU and CU are softwarized entities deployed at micro-datacentres, made up of COTS hardware. The O-RAN near-real time RIC is deployed at the same hierarchy as the DU given that the xApps will subscribe to information from various radio protocol layers while the management back-end of O-RAN is deployed in the central cloud. Certain 5G core components have been deployed in a more distributed fashion while others have been centralized depending on operational requirements. 

%O-RAN is committed to remain compliant with the 5G standardization carried out by the 3rd Generation Partnership Project (3GPP). With that in mind, the O-RAN community has its own set of technical specifications outlining the details of the architecture and working groups dedicated to specific areas of development~\cite{oranwg1archdesc}. 

While O-RAN adoption supports a flexible RAN deployment, it inadvertently introduces a new attack surface due to xApps being supplied by third-party vendors with various levels of trust. Moreover, given the decentralized and modular nature of 5G networks which will be developed as microservices, it is impractical to implement bulky, monolithic functionalities in a single VNF. 

This creates a demand for inter-xApp communications at a large scale, which requires secure and efficient execution of cross-component authorization and discovery in addition to permissions to access infrastructure functionality. Currently, both the 3rd Generation Partnership Project (3GPP)~\cite{3gpp23501} and the community implementation of OpenRAN lack the necessary security fundamentals to enable a scalable deployment.

Collaboration efforts~\cite{Commonwe39online,PublicRe8online} have started pursuing experimental work within O-RAN ~\cite{10114534770863480842,10114534588643466912,9647461,220513178} with focus on real-life measurements. Additionally, smaller scale testbeds have been constructed~\cite{9771908,9562627,9771860,200508374,220513178} using open-source solutions~\cite{bonati2020open,esmaeily2021small} with the aim of incorporating O-RAN compatible RAN nodes. However, none of these efforts addressed the inherent lack of security mechanisms inside the O-RAN framework.

To address this security gap, we propose the \textbf{xApp repository function (XRF)} framework, a microservice-based client-to-server augmentation to the O-RAN platform, which will provide authentication, authorization, and discovery mechanism for xApps in the O-RAN architecture analogous to the network functions repository function (NRF)~\cite{3gpp29510} in the 5GC. XRF provides a secure service discovery mechanism, facilitating the efficient interaction of xApps in a microservice architecture. In addition to maintaining xApp metadata, the XRF serves as an OpenAuthorization (OAuth) 2.0 server and distributes access tokens to xApps, enabling them to provide and consume services through secure API transactions.

%During its lifecycle, an xApp will be instantiated and register with the XRF. This registration is preceded by the mutual authentication of the xApp with the XRF, after which the former will send pre-determined metadata to the latter, to be used in advertising the service readiness of the xApp. 

%\textbf{The major contributions of this paper are} 
We complete the system design, implementation, and testing tasks of the XRF framework, which includes the modular server-side as a centralized entity and the lightweight client-side applications adjacent to xApps. %hat do not require any an internal integration with any existing O-RAN project codebase.
These entities are deployed on a highly-available (HA) Kubernetes cluster wrapped in a service mesh, hosted on an Openstack infrastructure, where various performance benchmarks demonstrate that the XRF framework scales efficiently in a multi-threaded environment under the microservice model of deployment.
    
The rest of the paper is structured as follows: Section~\ref{sec:relwork} discusses related work and specifications regarding OpenRAN and O-RAN security. In Section~\ref{sec:thrmdl}, we detail our threat model and framework requirements. Section~\ref{sec:frmwdsgn} describes the complete XRF system design and proposed O-RAN integration. The experimental setup followed by the performance benchmarks is presented in Section~\ref{sec:perfeval}. Finally, we conclude the paper with Section~\ref{sec:conc}.

%% file: relworks/texfile_v2.tex
In this section, we summarize previous work with prototype solutions for certain issues in the O-RAN architecture and the O-RAN Security Focus Group (SFG) efforts. 

\noindent \textbf{O-RAN security surveys and prototypes:} Several studies have been dedicated to mapping out the existing O-RAN attack surface and prototyping various security solutions independently from the specification efforts.

Authors of~\cite{220201032} have put together a primer, explaining the various workflows and communication interactions involved, as well as summarizing the work conducted by the O-RAN SFG ~\cite{oransfgprotspec,oransfgreqspec,oransfgthrmdl,oransfgpsectest}.
A systematic breakdown of the attack surface has been conducted for O-RAN in~\cite{220106080} by identifying the threat vectors of known physical and virtualization layer attacks towards entities in the architecture. While it does provide an overview, specific system design issues related to certain sub-components such as xApps have not yet been analyzed. 
Another analysis is provided in~\cite{220412227} where certain privacy and mitigation strategies for cloud environments are discussed in addition to discussing potential threats and actors. Authors conclude that the integration of O-RAN with specifically 5G will lead to an expanded attack surface.

%authors conclude that the security concerns will arise primarily from the interactions of O-RAN with 5G, acknowledging the fundamental issues which need to be addressed in such a large-scale integration. 

A technical study on open fronthaul interface security is conducted in~\cite{10114534654813470080}, where the mandatory security requirements for the management, control, user, and synchronization planes are presented. The authors then propose MACsec~\cite{8585421} as a candidate for securing the control user planes between the RAN entities where open interfaces are used. 

%Independently, the authors of~\cite{9604996} also propose MACsec as a preliminary method for addressing O-RAN fronthaul security. %However, the authors acknowledge that whilst using MACSec certain performance metrics need to be taken into consideration. 

\noindent \textbf{O-RAN security specifications:} The O-RAN development has been documented in a series of specifications by SFG.
%documented in a series of specification documents with a dedicated security focus group (SFG). %emphasizing security requirements. %divided into working groups (WG) given in Table.~\ref{tbl:oranwgs

\begin{comment}
\begin{table}[ht!]
\vspace{12pt}
\centering
\small\selectfont
\caption{O-RAN Working Groups}
\label{tbl:oranwgs}
\begin{tabular}{|p{0.03\textwidth}|p{0.255\textwidth}|}
\hline
WG1 & Use Cases and Overall Architecture \\
WG2 & Non-real-time RIC and A1 Interface \\
WG3 & Near-real-time RIC and E2 Interface \\
WG4 & Open Fronthaul Interfaces \\
WG5 & Open F1/W1/E1/X2/Xn Interface \\
WG6 & Cloudification and Orchestration \\
WG7 & White-box Hardware \\
WG8 & Stack Reference Design \\
WG9 & Open X-haul Transport \\
TIFG & Test and Integration Focus Group \\
SFG & Security Focus Group \\\hline
\end{tabular}
\end{table}
\end{comment}

%We are primarily interested in the work carried out by the SFG, which is internally divided into the security protocol~\cite{oransfgprotspec}, security requirements~\cite{oransfgreqspec}, test specifications~\cite{oransfgpsectest} and threat modelling and remediation analysis~\cite{oransfgthrmdl} sub-groups.

The O-RAN specific components extend the attack surface of the 5G system defined by 3GPP. This introduces additional security challenges and risks to consider while implementing these augmentations to the existing 3GPP architecture. In~\cite{oransfgprotspec}, the O-RAN SFG outlines the implementation requirements for security protocols used in O-RAN compliant interfaces. %O-RAN interfaces that handle sensitive operations (i.e., authentication, confidentiality, integrity) must support most up-to-date versions of the security protocols, disable insecure ciphers by default, and provide upgrade paths for future releases. %O-RAN echoes 3GPP's stance in suggesting cutting down on the number of selectable security parameter options to ensure interoperability. 

Furthermore, the group is working on specifying high-level security requirements and control mechanisms for O-RAN defined interfaces and network functions in~\cite{oransfgreqspec}. However, security requirements for entities of interest maintained by O-RAN are yet to be addressed. It is stated that interfaces must provide basic information security primitives such as confidentiality, integrity, authentication and access control for services and applications. 

Specifically for xApp and near-RT RIC security, key issues such as the deployment of malicious xApps or the compromise of existing xApps are identified in~\cite{oransfgxapps}. To address fundamental xApp security, network layer authentication solutions such as IPsec and TLS are proposed, followed by a high-level description of API authorization for xApps using the near-RT RIC as an authorization server. %is proposed in subsection (6.5.2) of~\cite{oransfgxapps}. 
The proposed solution lacks specificity in how it will handle xApp authentication and authorization at a large scale. Furthermore, it creates a reliance on the near-RT RIC as a trusted entity which is required to serve as an authorization server. Such a reliance will result in the implementation of critical security functionality next to the generic management modules, which will expose the former to internal attack vectors from the latter.

%Validation methods for testing %proper implementation 
%of security protocols, emulation of attacks, and validating possible mitigation strategies for O-RAN maintained components %interfaces, components, and services 
%are discussed in~\cite{oransfgpsectest}. %O-RAN SFG requires password-based authentication for every management plane protocol to be tested against brute force attacks. Network protocols and interfaces must be tested for robustness against fuzzing and distributed denial of service (DDoS) attacks. However, 
%Fundamental security testing and validation guidance such as software components security requirements and security testing for O-RAN components are discussed in~\cite{oransfgpsectest}. The document highlights the lack of security testing for elements such as the Near-RT RIC and xApps.

In~\cite{oransfgthrmdl}, roles and responsibilities are identified for maintaining and operating O-RAN systems which also includes prerequisites and assumptions to securely implement and run O-RAN defined components, interfaces, and protocols. A threat model is constructed by identifying the threat actors, describing the attack surface, identifying the vulnerabilities and assets, and listing affected components for each threat. %Finally, security principles and countermeasures are discussed for the identified threats. 

A recent study summarizes the work done by the O-RAN SFG~\cite{9728862} while considering certain 5G related concerns mentioned in~\cite{8894379}. Currently, it provides a concise overview of the perspective on security from the side of the O-RAN SFG. 
While the O-RAN SFG has taken the preliminary steps towards formulating the attack surface and identifying certain requirements, the majority of the surface remains unmapped. Additionally, no concrete frameworks have been implemented or proposed to address the lack of security primitives in inter-xApp communication. To the best of our knowledge, this paper is the first to design, implement and analyze a scalable authentication, authorization and discovery framework for xApps that addresses the fundamental system security requirements.

%% file: thrmdl/texfile.tex
The O-RAN architecture builds on top of the concepts and standards laid out by 3GPP~\cite{3gpp23501} for added functionality. However, new features such as virtualized functions, open interfaces, use of disaggregated COTS hardware, virtualization, and use of open-source code affect the 5G attack surface. Therefore, carefully defining the risks and scope considered for introducing a security element (i.e., the XRF) in the O-RAN architecture is crucial. 
Here we detail any relevant assumptions regarding the system with trusted and untrusted entities, along with the attacker model, objectives, and capabilities. %Finally, we describe the various attack vectors and attacker capabilities. 

\subsection{Assumptions}
\subsubsection*{\textbf{Operational and security requirements}} The deployment environment of the entities is assumed to be hardened against virtualization attacks aiming to break through sandbox environments (i.e., containers, virtual machines)~\cite{8792139,8056951,8523802,7562068}. Our approach does not protect against side-channel attacks from neighboring sandboxes, malicious code injections during instantiation, and privilege escalation into infrastructure managing entities such as container engines and hypervisors. 
%7562068

\subsubsection*{\textbf{Trusted entities}}
\begin{itemize}
    \item The RAN entities such as the RU, DU, and CU are equipped with the O-RAN E2 agent~\cite{orane2ap} for communicating with the near-RT RIC.
    \item \textit{Management and orchestration (MANO)} entity handling resource orchestration, VNF instantiation, and infrastructure management.
    \item O-RAN central deployment, which includes the \textit{service management and orchestration (SMO)} and the \textit{non-real-time RIC (non-RT RIC)} and interfaces to the near-RT RIC %(i.e., A1 and O1).
\end{itemize}

\subsubsection*{\textbf{Untrusted entities}}
\begin{itemize}
    \item \textit{MVNOs} that wish to instantiate an xApp. 
    \item \textit{near-RT RIC} of the O-RAN architecture which handles xApp registration and interfaces with the RAN nodes.
\end{itemize}

\subsection{Attacker Model}
\subsubsection*{\textbf{Actor}} Malicious xApps deployed through the MANO are seeking to subscribe to a RAN node for eavesdropping on or disrupting radio stack operations. 

\subsubsection*{\textbf{Objectives}} Take advantage of the lack of proper authentication and authorization mechanisms to compromise the security of other xApps and the near-RT RIC internal communications. A malicious xApp can passively target the RAN to extract or alter sensitive information (e.g., track user location, infer identification). Moreover, a malicious xAPP can affect a RAN node's performance or QoS requirements by exploiting resources and thereby affecting the availability of other services. As an active attacker, the malicious xApp can compromise the integrity of the data handled by legitimate xApps and perform unauthorized manipulation of the RAN through rogue signaling. 

\subsubsection*{\textbf{Capabilities}} The xApp has access to the information available in the near-RT RIC message router (RMR) before subscribing to other endpoints in the RMR. 

\subsection{Attack Vector}

\begin{figure}[t]
    \centering
    \includegraphics[width=\columnwidth,trim={9.2cm 6.4cm 9.7cm 5.9cm},clip]{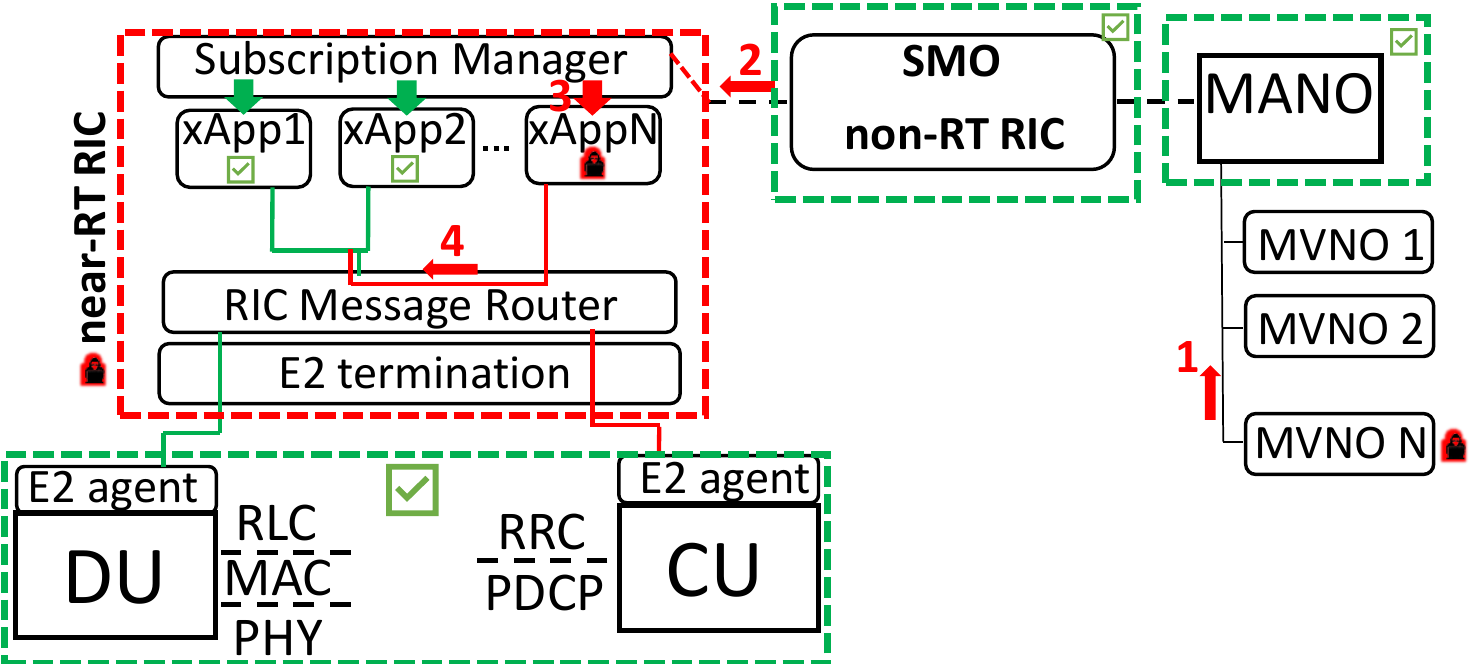}
    \caption{Attack vector propagation in the O-RAN system with a malicious xAppN and trusted/untrusted entities}
    \label{fig:thrmdl}
\end{figure}

Fig.~\ref{fig:thrmdl} illustrates the attack vector used to reach the objective while delineating the trusted and untrusted entities. In the first step, a malicious MVNO will instantiate an xApp by submitting a generic cloud template to the MANO. The xApp is marked for deployment at the near-RT RIC in step 2. Next, the xApp will register with the near-RT RIC, where the messages between the near-RT RIC entities are handled through the RMR. Once the xApp is ready, it will subscribe to the desired RAN stacks through the E2 termination in the near-RT RICs and the E2 agents on the RAN nodes~\cite{orane2gen}.

As a passive attacker, without any authentication or authorization method to control the access of xApps to the APIs of the critical entities, the malicious xAppN in Fig.~\ref{fig:thrmdl} can gain access to the information intended for the other xApps through the RMR. As an active attacker, xAppN can send rogue signals to the radio stack pretending to be xApp1 or xApp2, resulting in unwanted behavior from the RAN entities.

%% file: frmwdsgn/texfile_v2.tex
Our ultimate goal with this framework is to build a mechanism which will oversee the authentication, authorization, and discovery of xApps at a large scale. In the current microservice dominant implementation ecosystem, a given xApp will be required to consume services of other xApps for service-chaining their individual functionalities into a complete application. The XRF framework will facilitate this by servicing local xApps as a local metadata database and an OAuth2.0 server. It will distribute access tokens to service consuming xApps for them to securely access the APIs of service providing xApps.

In this section we first provide the reader with a brief overview of the relevant background on some of the fundamental design components that were used in the creation of the XRF framework. Next, details of the functional system design are provided followed by the message flow between server and client entities that were designed. We then dive into the non-functional details of the implementation in a production-grade environment and finally show a proposed integration of the framework into the O-RAN architecture to address the threat model described in Section~\ref{sec:thrmdl}.

\subsection{XRF Fundamentals}
Here we give a brief overview of the relevant background on some of the fundamental design components that were used in the creation of the XRF framework. 

\subsubsection{Open-Authorization 2.0}

\begin{figure}[t]
    \centering
    \includegraphics[width=\columnwidth,trim={0cm 0.5cm 0cm 0cm},clip]{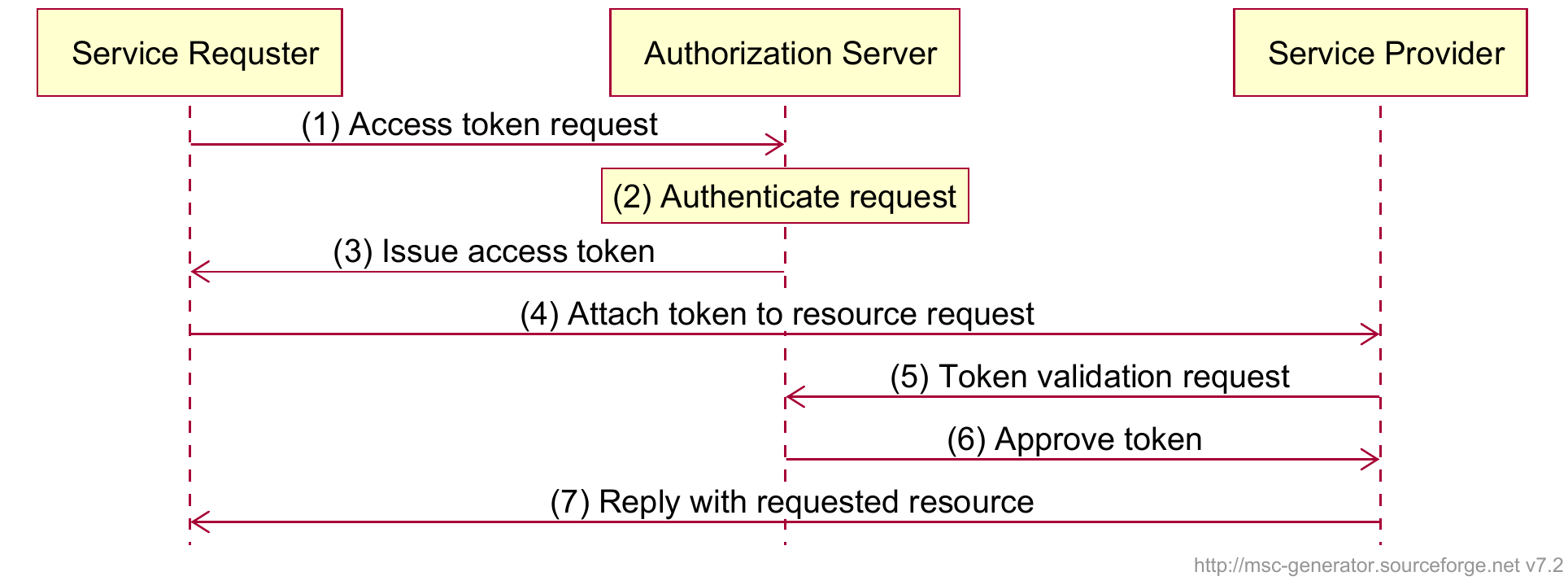}
     \caption{OAuth 2.0 Framework}
     \label{fig:backoauth2}
\end{figure}

OAuth 2.0 is a well-established standard where a service provider delegates authorization distribution rights to a trusted middle-man entity~\cite{RFC6749}. The main actors in the OAuth 2.0 framework are the service provider, service requester (client), service server and the authorization server. For the remainder of this paper, we will consider the service provider and the service server to be the same entity. The overall flow of the framework is given in Fig.~\ref{fig:backoauth2}. 
A service will start by requesting an access token from the authorization server for a target service. The server will authenticate this request and issue an access token. The service requester will use this token in accessing the API of the service provider. The provider will validate this request through the authorization server and, if the token is valid, will respond to the original request with the desired information. 

\subsubsection{Access Tokens}

\begin{figure}[t]
    \centering
    \begin{subfigure}[t!]{0.24\textwidth}
        \centering
        \includegraphics[width=\textwidth,trim={11.3cm 8.2cm 11.4cm 6.2cm},clip]{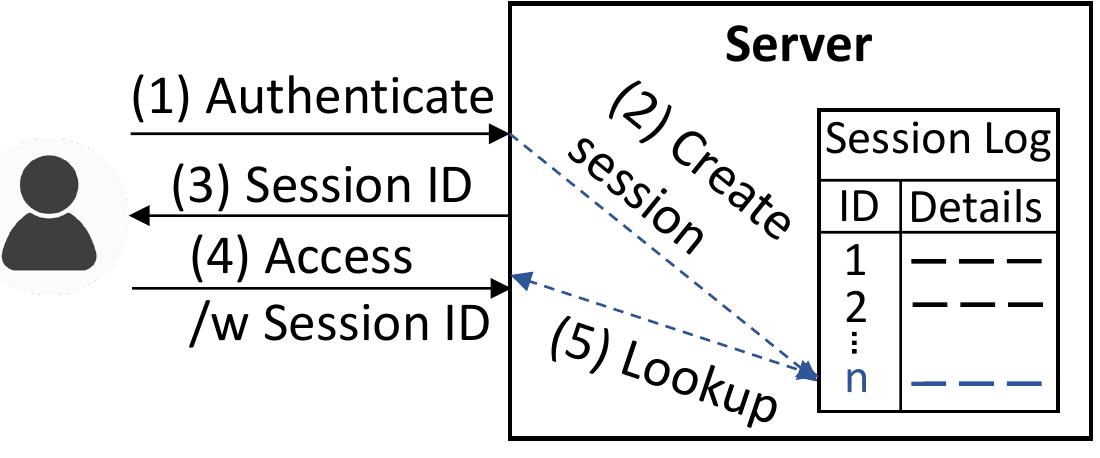}
        \caption[]%
        {{\small Session tokens}}    
        \label{fig:backsessionId}
    \end{subfigure}
    \hfill
    \begin{subfigure}[t!]{0.24\textwidth}  
        \centering 
        \includegraphics[width=\textwidth,trim={11.3cm 9.1cm 12cm 6.23cm},clip]{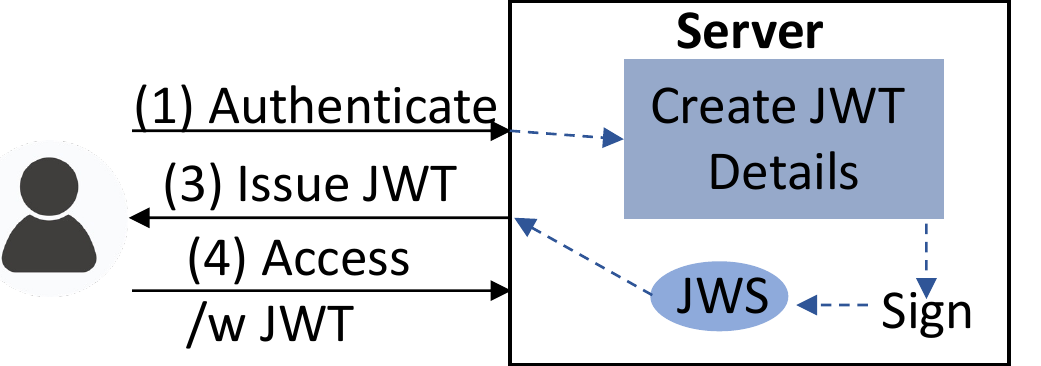}
        \caption[]%
        {{\small JSON Web Tokens}}    
        \label{fig:backjwt}
    \end{subfigure}
    \vskip\baselineskip
    \caption{Access token mechanisms}
    \label{fig:backtokens}
\end{figure}

The OAuth 2.0 framework~\cite{RFC6749} uses access tokens~\cite{RFC6750} to authorize service requests. These tokens are typically either session tokens or JSON web tokens (JWTs)~\cite{RFC7519}. The two frameworks are depicted in Fig.~\ref{fig:backtokens}.

For session tokens shown in Fig.~\ref{fig:backsessionId}, the server will maintain a database of sessions with all their details and associate each session with a unique ID. It will then issue this ID to the user and they can use it on subsequent access requests. This mechanism requires the server to maintain a log of sessions and perform look-ups to validate access requests. While it is a very popular method of authorization for web servers, in a microservice deployment, due to the number of services and volume of required lookups, it is not scalable.

JWTs, which are depicted in Fig.~\ref{fig:backjwt}, are more suitable for this purpose. In this case, all the access details for a particular service that would be required in any future request are parameterized in JSON format and signed by the server, forming a JSON web signature (JWS). This token is issued to the user and attached to any API call that will require it.

A signed JWT contains three elements, namely -- header, payload, and signature. The elements are Base64 encoded JSON key value pairs. The header contains general information such as signature algorithm (e.g., alg: RS256), unique key id (kid) associated with the token and media type (e.g., typ: JWT). Payload element contains the relevant user information such as unique identifier of the intended recipient or the service consumer (aud), expiration time of the JWT (exp), issuer (iss) of the JWT (e.g.,  authorization server), access type denoted as scope (e.g., read, write) and unique identifier of the token subject (sub). The issuer signs the header and the payload for the recipient to verify the authenticity of the token. %The key used for signing can be a symmetric or asymmetric key depending on the use case.

\subsubsection{Sidecar Proxies and Service Mesh}

\begin{figure}[b]
    \centering
    \includegraphics[width=\columnwidth,trim={6.1cm 6.55cm 12.5cm 6.6cm},clip]{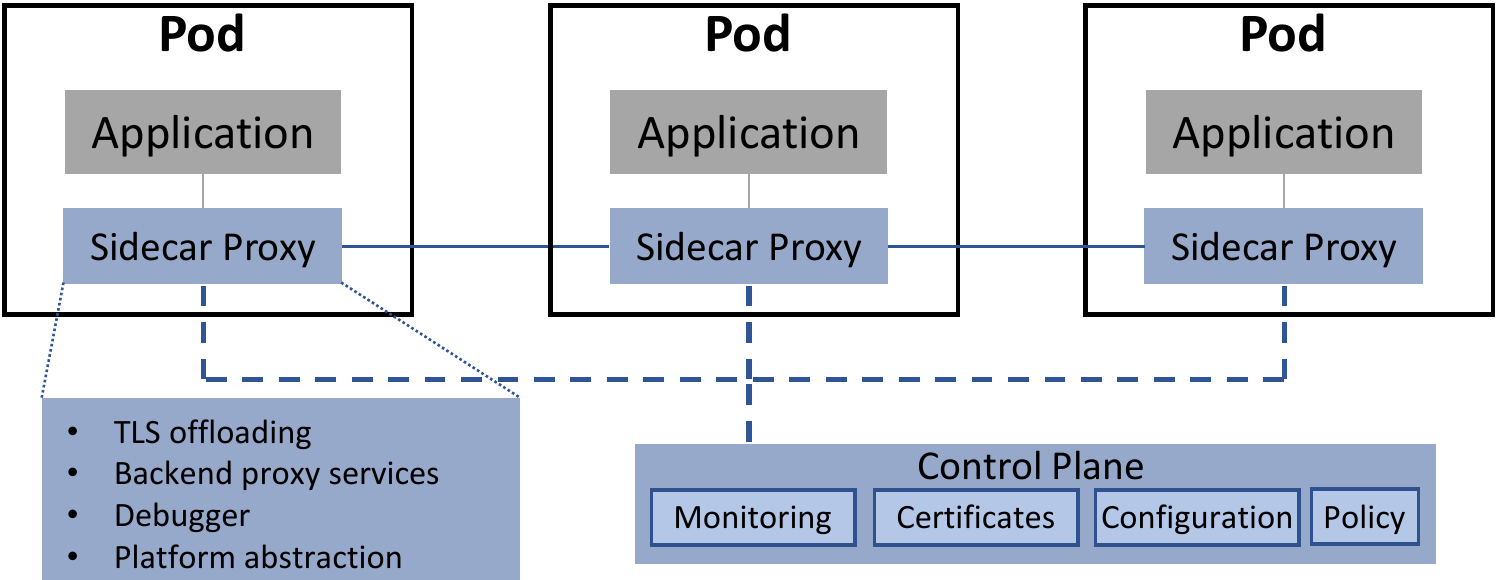}
     \caption{Sidecar proxy utilization in containerized environments over a service mesh}
     \label{fig:backscp}
\end{figure}

In the design of microservices for containerized environments, the concept of sidecar proxies (SCPs)~\cite{EdpriceMsft2022Jun} has become a popular choice for abstracting non-functional requirements from the main application. This is depicted in Fig.~\ref{fig:backscp}.

In a Kubernetes environment, where the smallest unit of deployment is a pod, a single pod can be comprised of multiple containers. One of these containers is the microservice application which is not exposed to the outside of the pod but supported by SCP containers. These SCPs provide non-functional security as well as other reverse proxy services to the primary application such as monitoring, load balancing and various other platform abstractions. The concept of SCPs has also been adopted in 3GPP standardization for 5GC VNFs in Rel.16 as a viable method of enabling inter-VNF communication~\cite{3gpp23501}.

These SCPs are connected by a service mesh~\cite{Whatisas49online} and share a common control plane in a Kubernetes deployment. The control plane can provide transport layer security (TLS) certificate management along with long-term monitoring, non-functional configurations and policies. 

\subsection{XRF Functional System Design} \label{sec:sysdsn}
%Before presenting the full overview of the XRF framework, we first go over the system design principles that it was built with. The functional and non-functional requirements are summarized in Table\ref{tbl:sysreqs}. 

\begin{figure}[t]
    \centering
    \includegraphics[width=\columnwidth,trim={11.3cm 5.5cm 11.5cm 3.4cm},clip]{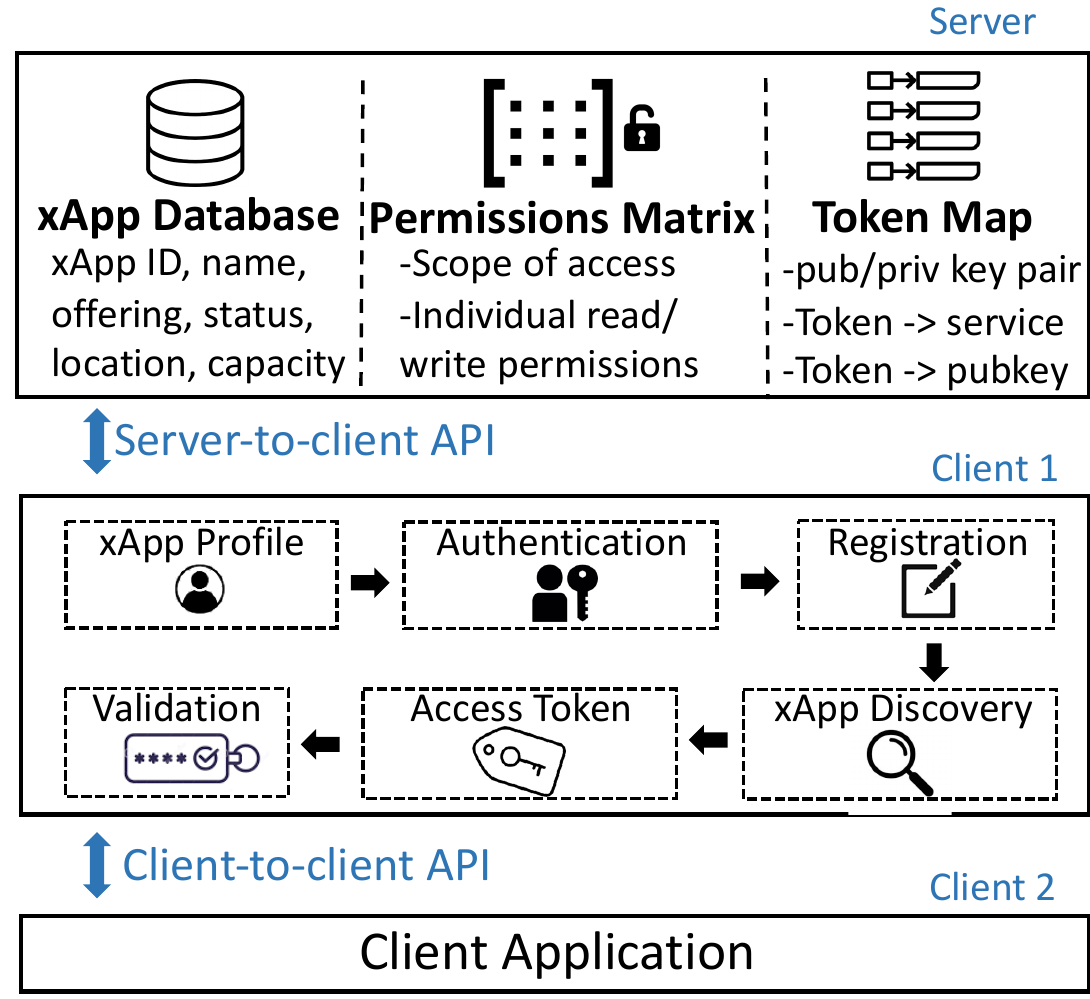}
     \caption{XRF functional system design overview with server and client internals}
     \label{fig:xrfsysoverview}
\end{figure}

The functional entities of the XRF system are presented in Fig.~\ref{fig:xrfsysoverview}. The framework is composed of two standalone entities, which are the XRF client module and XRF server. The high-level client flow and server internals are depicted in Fig.~\ref{fig:xrfsysoverview}.

To enable server-to-client and client-to-client communication, representational state transfer (REST) APIs have been constructed on both the server and the client. These are described in Table~\ref{tbl:xrfsysapi}. The XRF server runs a multi-threaded HTTP server with seven distinct endpoint handlers to offer a concise collection of request and response loops necessary to handle the lifecycle operations of a client.

\begin{comment}
\begin{figure}[b!]
    \centering
    \includegraphics[width=\columnwidth,trim={8.5cm 9.9cm 10.4cm 3cm},clip]{figures/xrfapioverview.pdf}
     \caption{XRF server and client APIs}
     \label{fig:xrfsysapi}
\end{figure}
\end{comment}

\begin{table}[b!]
\vspace{12pt}
\centering
\small\selectfont
\caption{XRF server and client APIs} 
\label{tbl:xrfsysapi}
  \begin{tabular}{p{0.05\textwidth}|p{0.39\textwidth}}
    \multirow{1}{*}{\textbf{System}} & \multirow{1}{*}{\textbf{API}} \\
    \hline
    \hline
    \multirow{7}{*}{Server} & \textit{POST}/InitialAuthentication(Challenge) $\rightarrow$ Counter \\
                            & \textit{PUT}/RegistrationHandler(xAppProfile) $\rightarrow$ HTTP::OK \\
                            & \textit{PUT}/ProfileUpdateHandler(xAppData) $\rightarrow$ HTTP::OK \\
                            & \textit{GET}/xAppDiscoveryHandler(targetProfile) $\rightarrow$ List \\
                            & \textit{POST}/AccessTokenRequest(Func, Loc) $\rightarrow$ JWT \\
                            & \textit{POST}/TokenIntropsection(JWT) $\rightarrow$ HTTP::OK \\
                            & \textit{GET}/JWKSRequestHandler(keyID) $\rightarrow$ token pubkey \\
    \hline
    \multirow{2}{*}{Client} & \textit{PUT}/ProfileUpdate(xAppData) $\rightarrow$ HTTP::OK \\
                            & \textit{X}/ServiceRequestAPI -H Bearer "JWT" $\rightarrow$ desired \\
                              
  \end{tabular}
\end{table}

When it is first instantiated, the XRF client will create a metadata profile for the xApp. The profile parameters are given in Table~\ref{tbl:xappprofile}.  They are the instance identifier, an operator-chosen descriptive name, a standardized service offering code, current status, the physical location, and finally service load indicating how many clients a given xApp is currently serving. 

After the local creation of the xApp profile, the XRF client will start a sequence of lifecycle operations until it becomes service ready. First, it will create an authentication challenge for the \textit{InitialAuthentication} endpoint of the server and receive a counter-challenge for mutual authentication. It will then transform the xApp profile metadata into JSON format and send it to the \textit{RegistrationHandler} endpoint of the XRF server where it will be stored in the xApp database.

The XRF server will maintain the metadata of clients using an in-memory key-value storage. It will keep track of the service consumers of any registered client and in the event that the profile of a provider changes, it will notify the relevant XRF clients of xApps through their \textit{ProfileUpdate} endpoint.

\begin{table}[t]
\vspace{12pt}
\centering
\small\selectfont
\caption{xApp profile metadata} 
\label{tbl:xappprofile}
  \begin{tabular}{r|p{0.29\textwidth}}
    \multirow{1}{*}{\textbf{Data}} & \multirow{1}{*}{\textbf{Description}} \\
    \hline
    \hline
    \multirow{1}{*}{xAppInstanceID} & a universally unique identifier (UUID) \\ \hline
    \multirow{1}{*}{xAppInstanceName} & human-readable name for the instance \\ \hline
    \multirow{1}{*}{xAppOffering} & \textit{ServiceRequestAPI} endpoints offered \\ \hline
    \multirow{1}{*}{xAppStatus} & availability status of the xApp \\ \hline
    \multirow{1}{*}{xAppLocation} & physical deployment location  \\ \hline
    \multirow{1}{*}{xAppLoad} & number of serviced xApps  \\ \hline
  \end{tabular}
\end{table}

%At this point, the xApp has been fully registered with the server and is ready to carry out the relevant functionality that it was created to fulfil in the microservice chain. To initiate or join the service-chain, the service consumer xApp will need to access the services provided by a service provider xApp.

In the steps towards joining a microservice chain, the client will craft a discovery request to the XRF server \textit{xAppDiscoveryHandler} endpoint for a target client that is advertising the desired \textit{xAppOffering} code in the profile.  To respond to a discovery request, the server will query the internal xApp database and respond with a list of eligible profiles. 

The client will then formulate a request for the \textit{AccessTokenRequest} endpoint of the server to retrieve an access token for consuming services from a target provider. This token can be attached in the header of API calls whenever required. 

With token distribution, the XRF server performs access enforcement on target clients where the scope of a token is determined by an internal permissions matrix populated using the value of the \textit{xAppOffering} parameter from Table~\ref{tbl:xappprofile}. 

The \textit{xAppOffering} advertises a standardized and globally comprehensible xApp functionality while also indicating the predetermined access rights allowed at the service endpoints of a client. If an incoming access token request conflicts with the recorded entry (e.g. write request to a read-only endpoint), the server will reject this request.

Furthermore, the server will use the permissions matrix to resolve any conflicting access rights so that no more than one consumer is granted critical access to specific services of the same provider at the same time. For instance, two consumers might concurrently try to write to the same data in a provider, which can result in service conflicts. %Such a case will be prevented using the permissions matrix.

The distribution of access tokens makes the XRF server an OAuth 2.0 authorization server where the XRF client modules have a service consumer and provider back-and-forth flow. The tokens in our framework are JWTs which are bound to a unique RSA key pair where the private key is used to sign the token and the public key is used to verify them.

Any client can perform token validation on a service request through either a self-contained validation using the \textit{JWKSRequestHandler}, where the token is validated internally through the server, or assisted validation where the token is forwarded to the \textit{TokenIntropsection} of the server. %Upon successful validation of the token, the consumer and provider can resume communication.

\subsection{XRF Server-Client Information Flow}
\begin{figure}[t]
    \centering
    \includegraphics[width=\columnwidth,trim={0cm 0.5cm 0cm 0cm},clip]{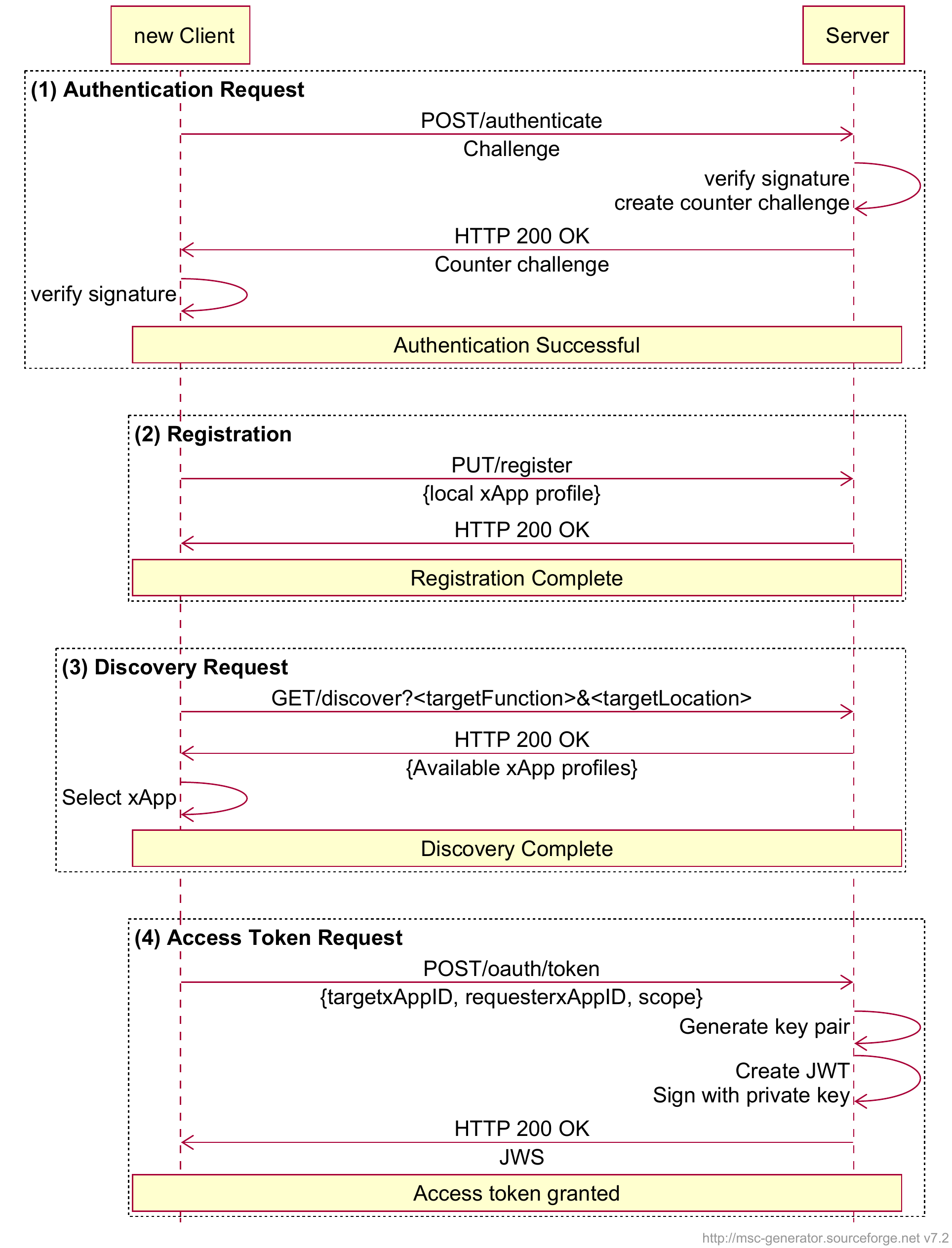}
    \caption{Preliminary operation flow between XRF client and server showing (1) initial authentication, (2) registration, (3) discovery request, (4) access token request}
    \label{fig:xrfc_s_prelim}
\end{figure}

The first set of messages exchanged between the XRF client and the server are shown in Fig.~\ref{fig:xrfc_s_prelim} starting with the instantiation of the client until it becomes service ready after obtaining an access token for consuming services from other client modules. 

%The first thing that the client will do after being instantiated is authenticate (1) with the server according to the algorithm described in \cite{stallingsbook} using OpenSSL cryptographic suite.
The first step the client performs after creating a profile is to authenticate (1) with the server through public key authentication. %Client generates a random number $m$, creates hash $H(m)$, and signs it with its private key $PR_{XRF_c}$. Then the signature is appended with the random number and encrypted with the server public key $PU_{XRF_s}$ so that only the server can read the contents of the challenge. The challenge is then Base64 encoded and sent as a JSON string. Upon receiving the challenge, server decrypts it with the corresponding private key $PR_{XRF_s}$ and verifies the signature using the client public key $PU_{XRF_c}$. If the signature is verified, server signs $m$ with its own private key $PR_{XRF_s}$ and encrypts it with the public key $PU_{XRF_c}$ of the client and sends it back. Similarly, the client decrypts the contents and authenticates the server by verifying the signature.     

Once the authentication is successful, the client will register (2) by sending the required metadata in a PUT request to the server. This includes the parameters described in Table~\ref{tbl:xappprofile}.

The client has now concluded the preliminary operations and is ready to become part of the microservice-chain. To proceed, it will query the server for a discovery (3) request for a target xApp. This request will include the desired \textit{xAppOffering} and \textit{xAppLocation} from Table~\ref{tbl:xappprofile} that the client is expecting from a \textit{ServiceRequestAPI}. The server will respond with the available list of profiles and the client will select the candidate with the lowest current \textit{xAppLoad} among the received profiles.

In order to consume the API of a target provider, the client will create an access token request (4) with the UUID of this target, its own UUID and the scope of the access that it wants from the chosen client. If the requested scope is allowed in the permissions matrix of the server as described in Section~\ref{sec:sysdsn}, a JWT will be created using a new public-private RSA key pair. The server will sign the JWT with the private key to create a JWS and deliver it to the client.

\begin{figure}[t]
    \centering
    \includegraphics[width=\columnwidth,trim={0cm 0.5cm 0cm 0cm},clip]{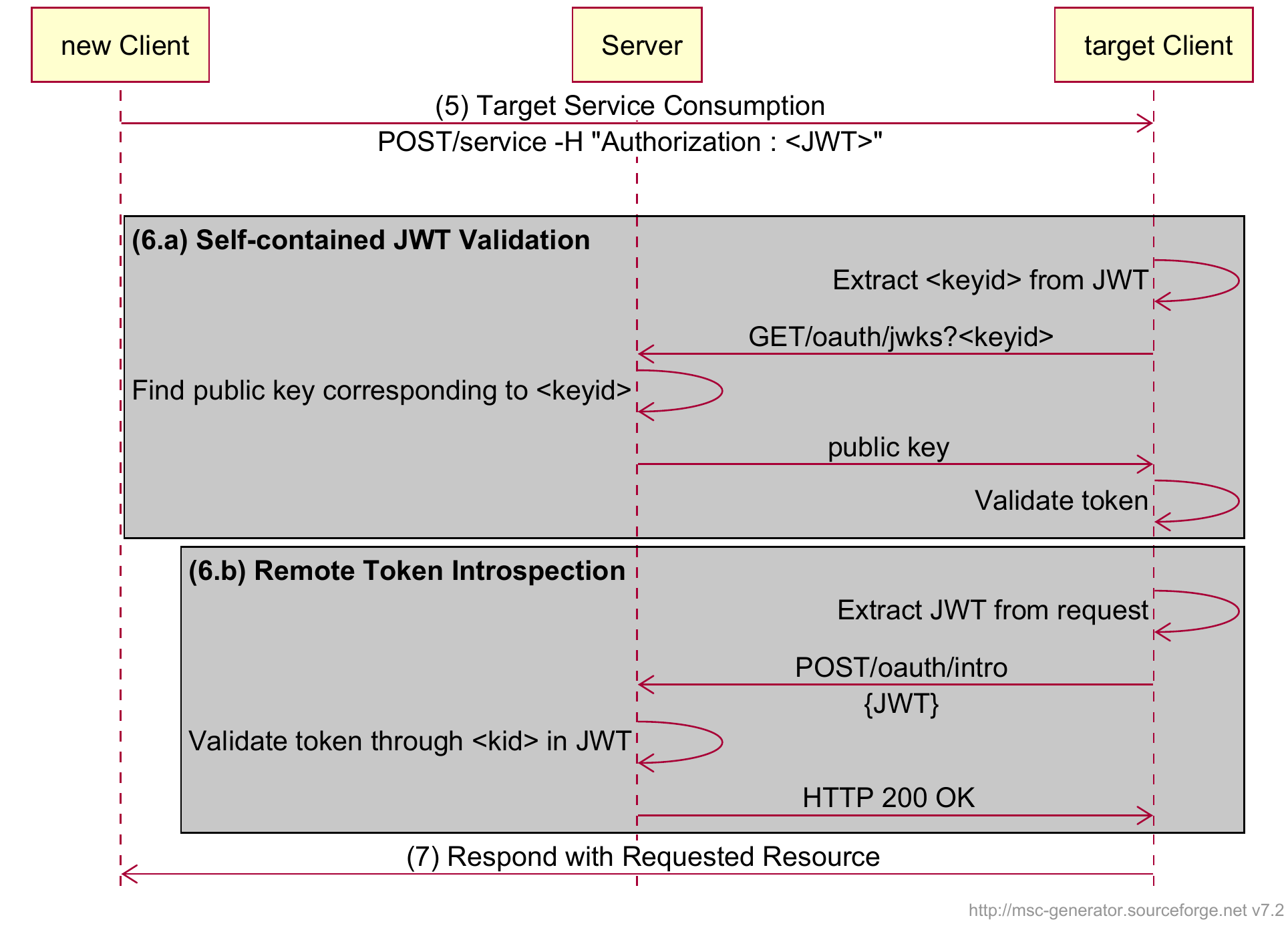}
    \caption{(5) Service consumption from a target client of another xApp using a token bearer with (6.a) self-contained JWT validation and (6.b) remote token introspection}
    \label{fig:xrfc_s_cons}
\end{figure}

Having received the access token, the client can now request service consumption from the target XRF client of another xApp, which is depicted in Fig.~\ref{fig:xrfc_s_cons}. %In this scenario, the "target client" has already authenticated and registered with the XRF while the "new client" has only just conclude the steps in Fig.~\ref{fig:xrfc_s_prelim}. The steps are described below.

\begin{figure}[b]
    \centering
    \begin{subfigure}[h!]{0.44\textwidth}
        \centering
        \includegraphics[width=\textwidth,trim={4cm 9.3cm 5.7cm 3.6cm},clip]{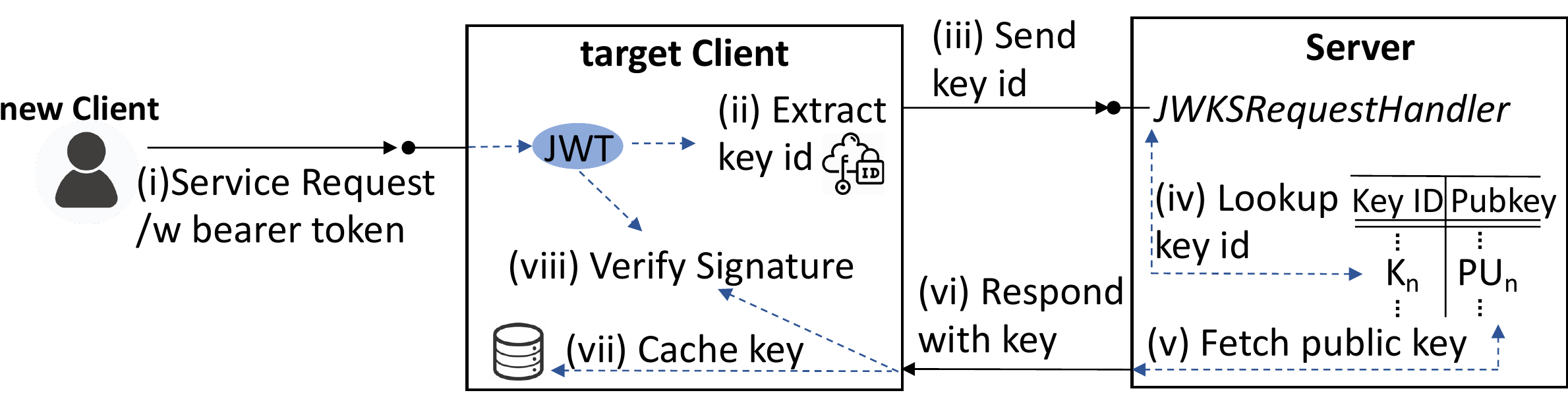}
        \caption[]%
        {{\small Self-contained JWT validation}}    
        \label{fig:jwksvalid}
    \end{subfigure}
    \hfill
    \vskip\baselineskip
    \begin{subfigure}[h!]{0.44\textwidth}  
        \centering 
        \includegraphics[width=\textwidth,trim={4cm 9.3cm 6.6cm 4.6cm},clip]{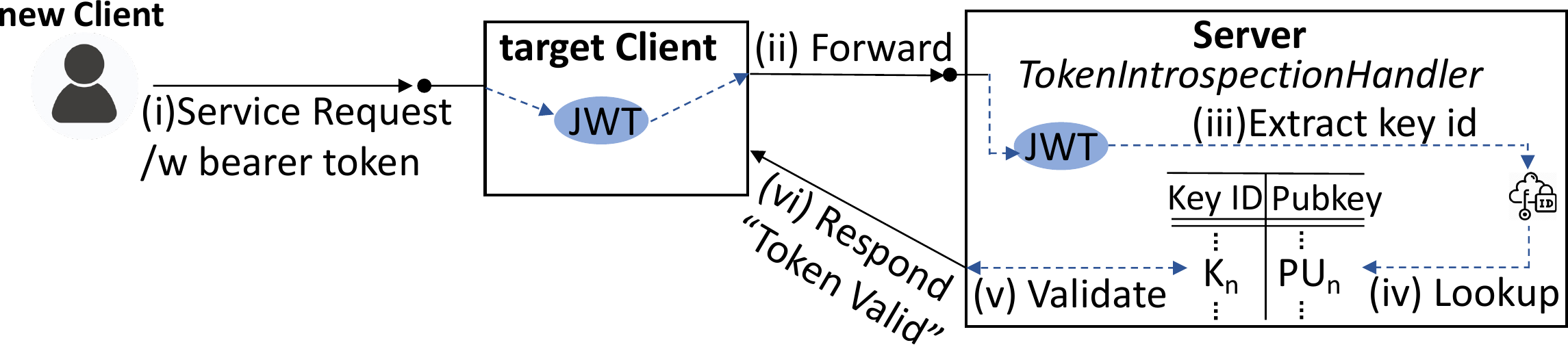}
        \caption[]%
        {{\small Remote JWT introspection with server}}    
        \label{fig:introvalid}
    \end{subfigure}
    \vskip\baselineskip
    \caption{Implemented access token validation mechanisms}
    \label{fig:tokenval}
\end{figure}

The token obtained in step (4) is attached to the header of the API calls (5) destined for the XRF client of the chosen xApp. Upon receiving the API request, the client can utilize two methods of validating the token (6).  These are depicted in Fig.~\ref{fig:tokenval}.

\textbf{(6.a) Self-contained JWT validation} is shown in Fig.~\ref{fig:jwksvalid}. After receiving the service request (i) the target client will extract the key id (ii) from the header of the JWT and query the JSON web key set (JWKS) endpoint \textit{JWKSRequestHandler} in Table~\ref{tbl:xrfsysapi} of the server with this value (iii). This key id will be internally mapped to the public key of the private key used in signing the JWT (iv-v). The XRF server will respond with this public key (vi) and the target client will cache it (vii) for future API calls with the same key id and then proceed to validate the token (viii). The advantage of this approach is that, after the first API call, the target client can perform validation on the same key id independently from the server. However, the approach expects token validation functionality to be implemented on the XRF client.

\textbf{(6.b) Remote token introspection} is shown in Fig.~\ref{fig:introvalid}. The target client will extract the JWT from the header of the HTTP request and forward it entirely to the \textit{TokenIntrospection} endpoint of the server (ii) where the latter will validate it in the same way (iii-v) as the client in (6.a) and respond with the "token valid" message (vi). 

Once the token has been validated, the target client can now respond (7) to the new client with the requested resource. 

\subsection{XRF Implementation}
The implementation of the XRF server and client applications is done in C++17 in Ubuntu 20.04. Both applications are running an HTTP server with REST API endpoints crafted using the Pistache library~\cite{GitHubpi38online}. Pistache is a mature C++ API compatible with OpenAPI~\cite{Hellofro97online}. To make the API accessible, we have first prepared the human-readable YAML files with the OpenAPI 3.0.0 specification and later used the OpenAPI generator to create template APIs that were populated with our custom endpoint handlers.

For handling access tokens, we used a JWT library~\cite{GitHubar4online} which uses OpenSSL as a primary dependency. It provides encoding and decoding of access tokens which can be used for creation, signing and validation. 

\begin{figure}[t]
    \centering
    \includegraphics[width=\columnwidth,trim={3cm 2.4cm 9cm 3.5cm},clip]{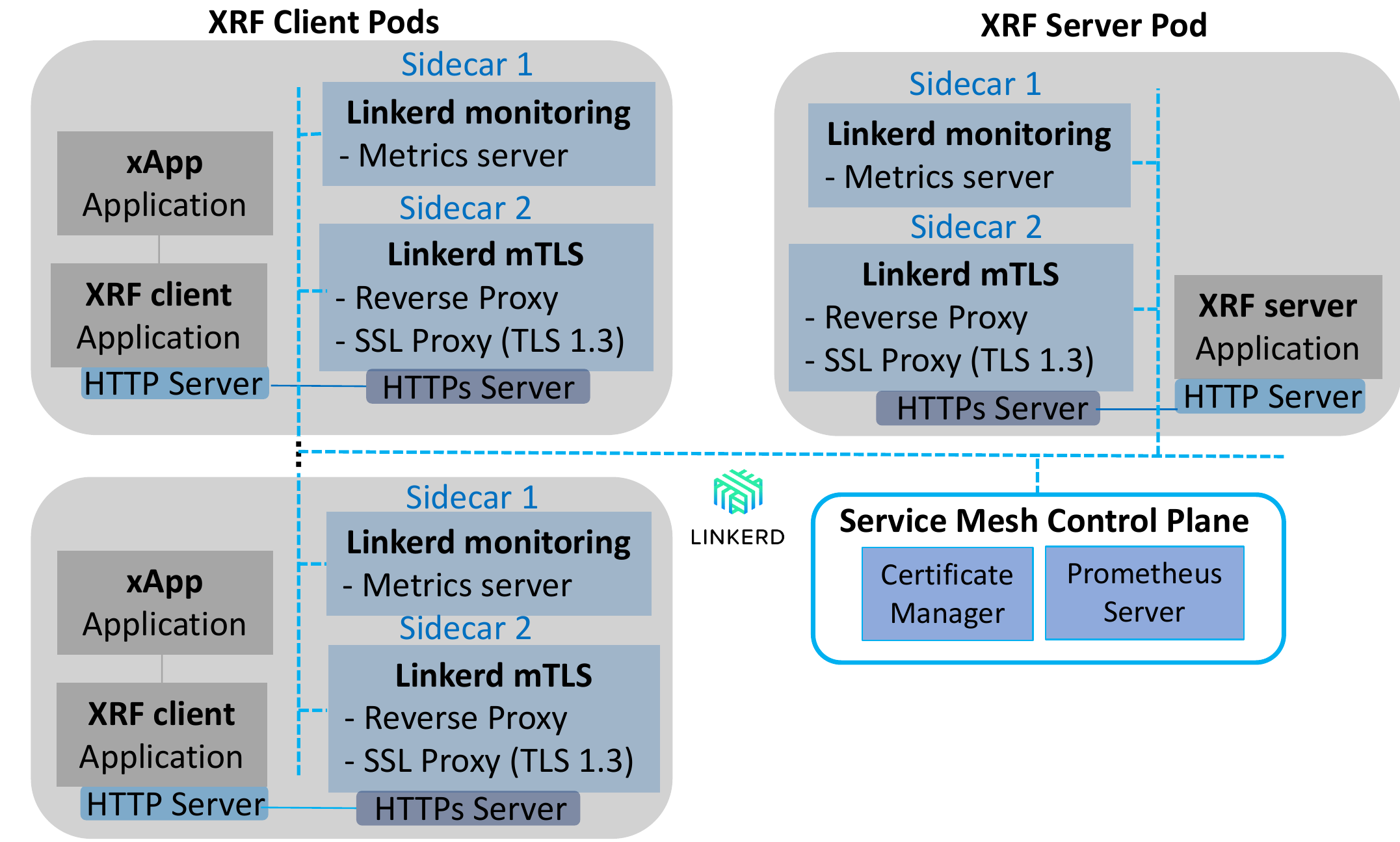}
    \caption{Full Kubernetes deployment model with application containers and auxiliary functionality providing SCPs connected with a Linkerd service mesh}
    \label{fig:k8details}
\end{figure}

The Kubernetes deployment model is given in Fig.~\ref{fig:k8details}. It shows the deployment of the XRF client and XRF server pods. The former is made up of four containers, where the xApp application conceptually sits in the pod back-end and only has communication with the XRF client container. Depending on the design, this can be inter-process communication (IPC) or direct integration of the client libraries into the xApp application. The client container is running an HTTP server which is not exposed to the network but is connected to the Linkerd mutual TLS (mTLS)~\cite{Whatisas49online} SCP, which is serving as a back-end reverse proxy. The main use of this SCP is to serve as a secure socket layer (SSL) proxy, which encrypts the HTTP traffic from the client module with TLS 1.3 and exposes an HTTPs server to the network using self-signed certificates. 

The SCPs are connected using Linkerd~\cite{Whatisas49online}, a lightweight service mesh solution which provides a centralized monitoring server and certificate manager that rotates TLS certificates over time. To take advantage of the monitoring tool, we also deploy the Linkerd monitoring SCP which connects to the central Prometheus server to enable long-term metrics gathering.

The same implementation is mirrored on the XRF server pod where the server application is separate from the Linkerd SCPs. This model of deployment ensures that the microservice application is not burdened with implementing any non-functional security features. 

\begin{figure}[t]
    \centering
    \includegraphics[width=\columnwidth,trim={6.7cm 8.1cm 11cm 6.4cm},clip]{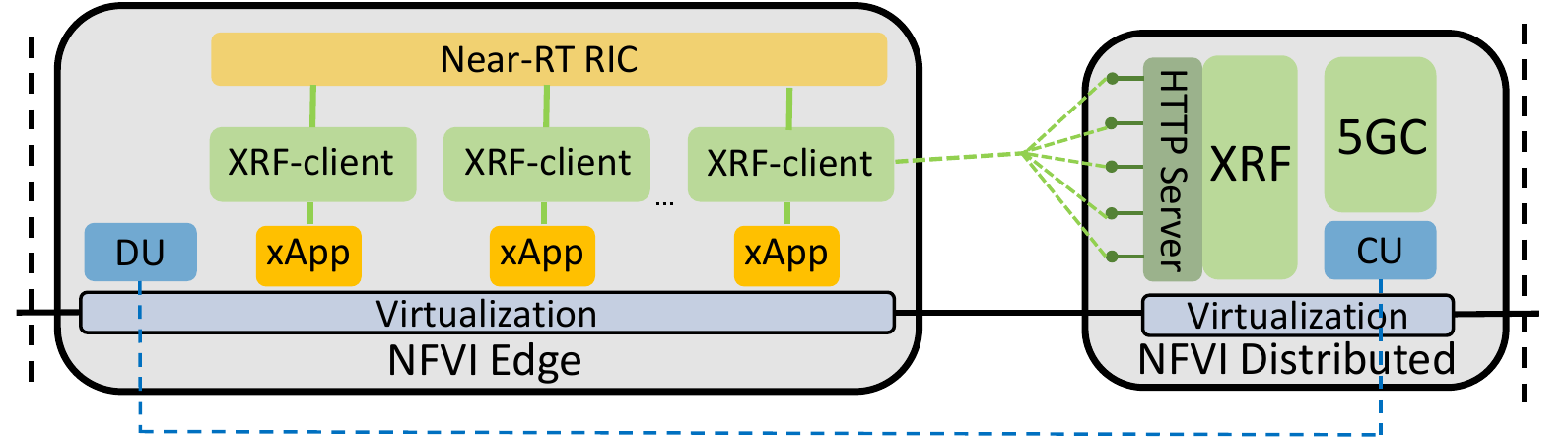}
    \caption{Proposed XRF integration into the O-RAN ecosystem}
    \label{fig:xrfintegration}
\end{figure}

%%Removing this subsection as it doesn't seem necessary and looks too short for a standalone subsection
%\subsection{Proposed XRF Integration with O-RAN} 
 A proposed integration of the XRF framework into the O-RAN and 5G architecture is presented in Fig.~\ref{fig:xrfintegration}. The XRF server functionality is exposed to the XRF clients through REST API endpoints. The client module is a microservice application independent from the primary xApp. It plays the role of a middle-man entity between the xApp and the near-RT RIC, performing the operations described in Section~\ref{sec:sysdsn}. The information routing through the RMR, mentioned in Section~\ref{sec:thrmdl}, is bounced through the XRF client pod (as shown in Fig.~\ref{fig:k8details}) before reaching the xApps.
 
 This procedure requires the xApp to supply the XRF client with the metadata in Table~\ref{tbl:xappprofile} when it is first instantiated. From this point forward, in any xApp-to-xApp or xApp-to-near-RT RIC communication, the XRF client will behave as a proxy to provide functional security with the flow illustrated in Fig.~\ref{fig:xrfsysoverview}.
 
As an entity that oversees multiple edge domains at the same time, the XRF server can be deployed in the distributed cloud. Depending on the operational requirements it can be placed in the edge network for lower latency or further centralized for a broader management scope.

%% file: perfeval/texfile.tex
In this section, we present our experimental setup for testing with the XRF framework, followed by our findings from various performance benchmarks on the XRF client and server. First, a set of operational throughput and latency experiments are carried out in a multi-threaded environment with varying wall and CPU times. Next, micro-benchmarks are performed over the individual modules in the XRF framework.

\subsection{Experimental Setup} \label{sec:expsetup}
\begin{figure}[t]
    \centering
    \includegraphics[width=\columnwidth,trim={9.3cm 6.6cm 12.5cm, 5.8cm},clip]{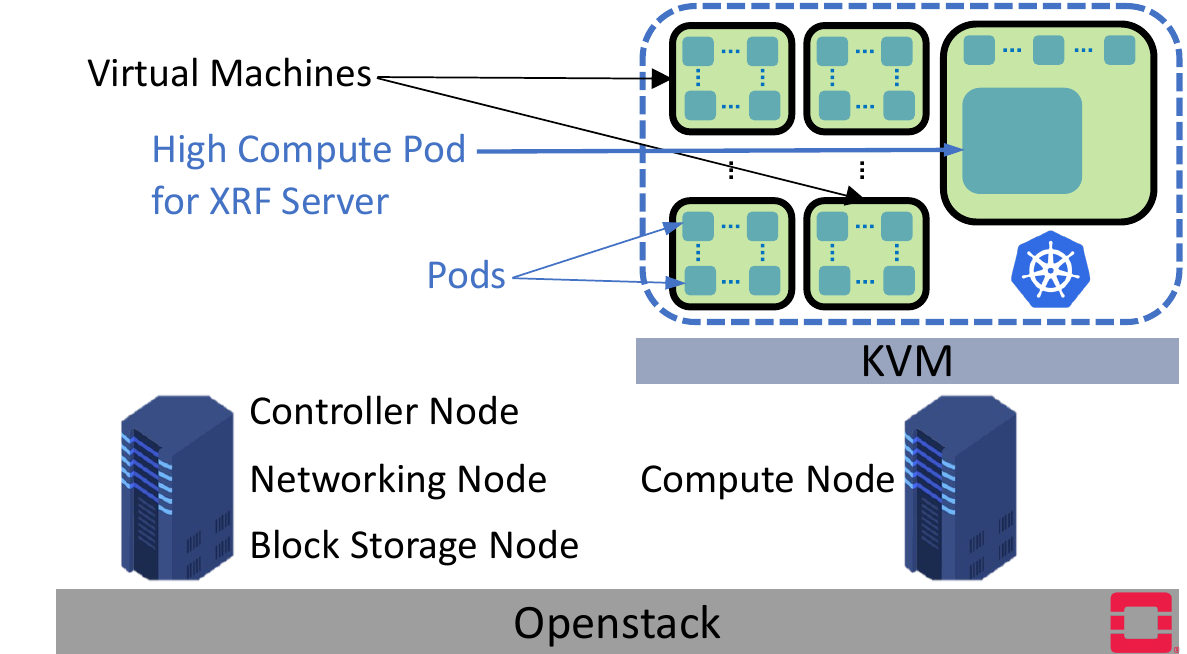}
    \caption{Testbed with Openstack virtual machines hosting a Kubernetes cluster}
    \label{fig:expsetup}
\end{figure}

Our testing environment, depicted in Fig.~\ref{fig:expsetup}, was used to deploy XRF client and server connections at a large scale where each client is instantiated inside a separate container.

We have two servers which are running a small-scale Openstack infrastructure. One of the nodes is serving as a controller, networking and block storage node while the other serves as a compute node with the KVM hypervisor. The nodes used are two Precision 7920 Tower servers with
%\begin{itemize}
  %  \item 
  2 x Intel Xeon Gold 5218R 2.1GHz CPUs,
    %\item 
    512GB RAM,
    %\item 
    1TB disk space, and running
    %\item 
    Ubuntu 20.04.
%\end{itemize}

The compute node is running a total of 11 virtual machines (VMs) for the HA-Kubernetes cluster constructed through the Ranchers Kubernetes Engine (RKE)~\cite{rkeengine}. Three VMs are control nodes and the remaining eight are workers, one of which is large (L-worker) and dedicated to the XRF server enabling all the vCPUs to prioritize server threads. The others are small workers (S-worker). The flavors are given in Table~\ref{tbl:kubclusconf}. 

\begin{table}[ht!]
\vspace{12pt}
\centering
\small\selectfont
\caption{HA Kubernetes cluster VM flavors}
\label{tbl:kubclusconf}
%\renewcommand{\arraystretch}{1.1} % for the vertical padding
%\resizebox{\textwidth}{!}{
\begin{tabular}{|p{0.2\columnwidth}|p{0.15\columnwidth}|p{0.1\columnwidth}|p{0.1\columnwidth}|p{0.1\columnwidth}|}
\hline
Node & Instances & vCPUs & RAM (GB) & Disk (GB) \\\hline
Control  & 3 & 2 & 8 & 40 \\\hline
S-worker  & 7 & 6 & 16 & 80 \\\hline
L-worker  & 1 & 20 & 200 & 80 \\\hline
\end{tabular}
%}
\end{table}

\subsection{Results and Discussion}
\begin{figure}[h!]
    \centering
    \begin{subfigure}[b]{0.475\columnwidth}
        \centering
        \includegraphics[width=\textwidth,trim={0cm 0cm 24.2cm, 0cm},clip]{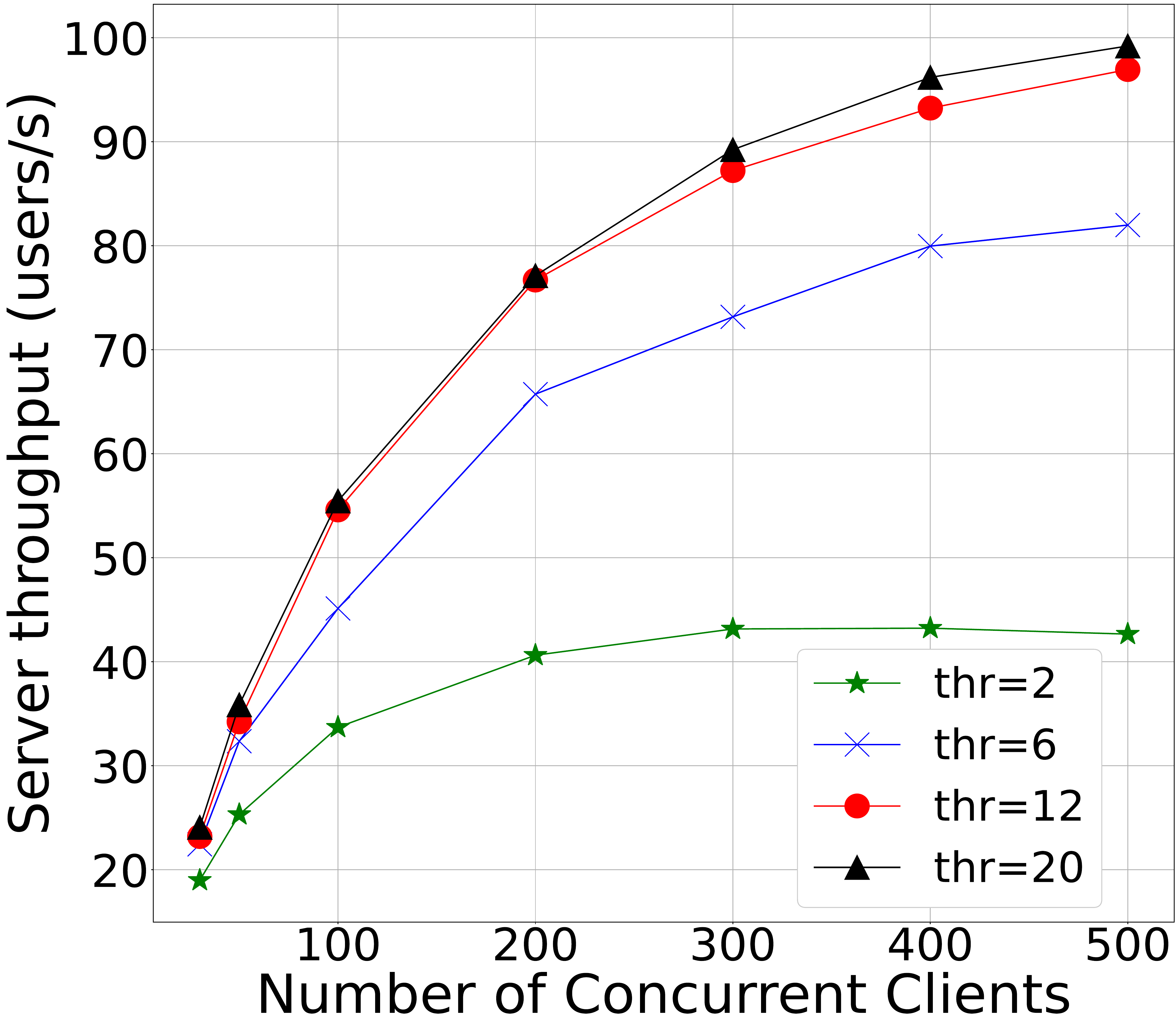}
        \caption[]%
        {{\small Server end-to-end throughput}}    
        \label{fig:thr_throu}
    \end{subfigure}
    \hfill
    \begin{subfigure}[b]{0.475\columnwidth}  
        \centering 
        \includegraphics[width=\textwidth,trim={0cm 0cm 25cm, 0cm},clip]{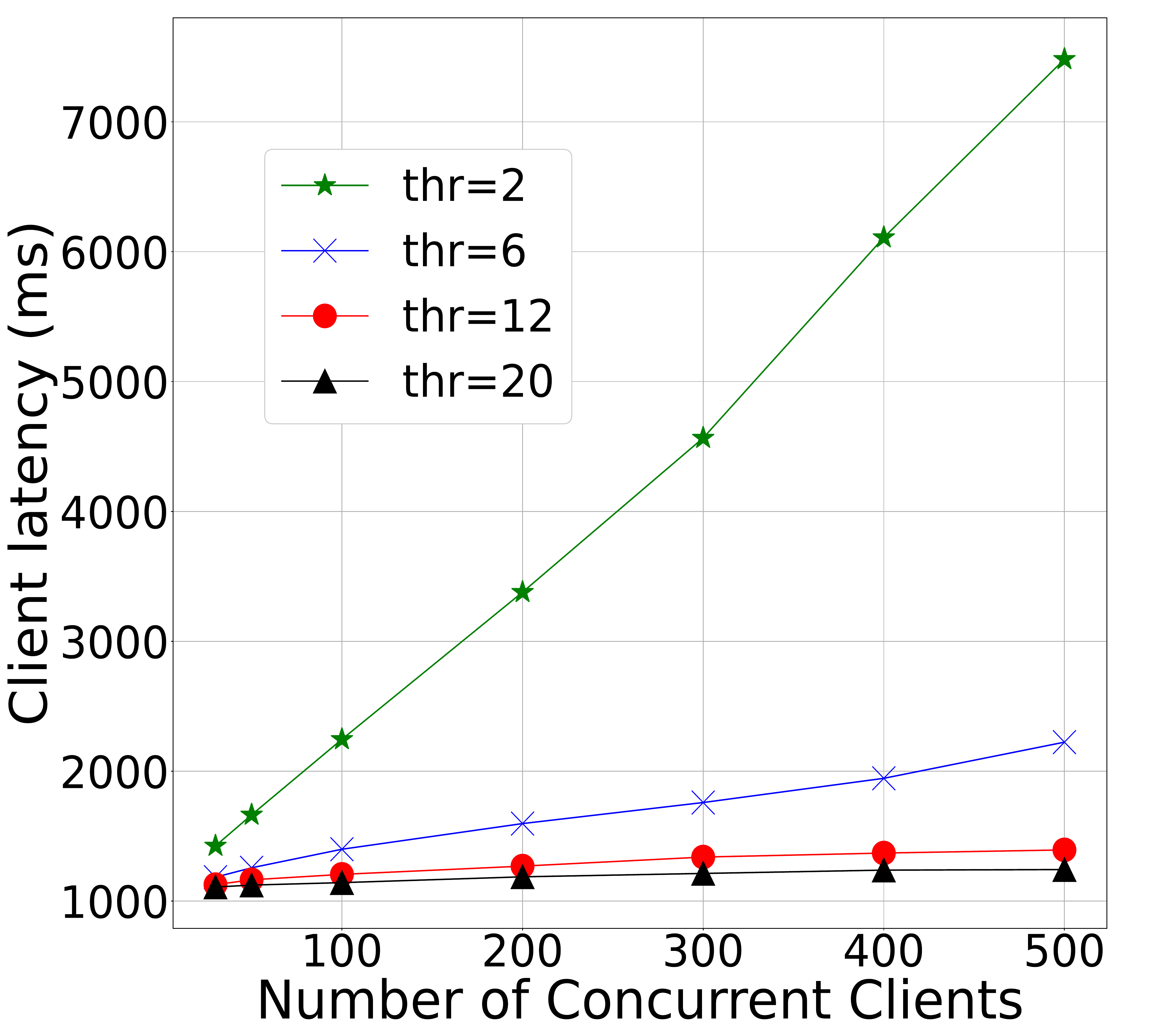}
        \caption[]%
        {{\small Client end-to-end latency}}    
        \label{fig:thr_lat}
    \end{subfigure}
    \vskip\baselineskip
    \begin{subfigure}[b]{0.475\columnwidth}   
        \centering 
        \includegraphics[width=\textwidth,trim={0cm 0cm 21cm, 0cm},clip]{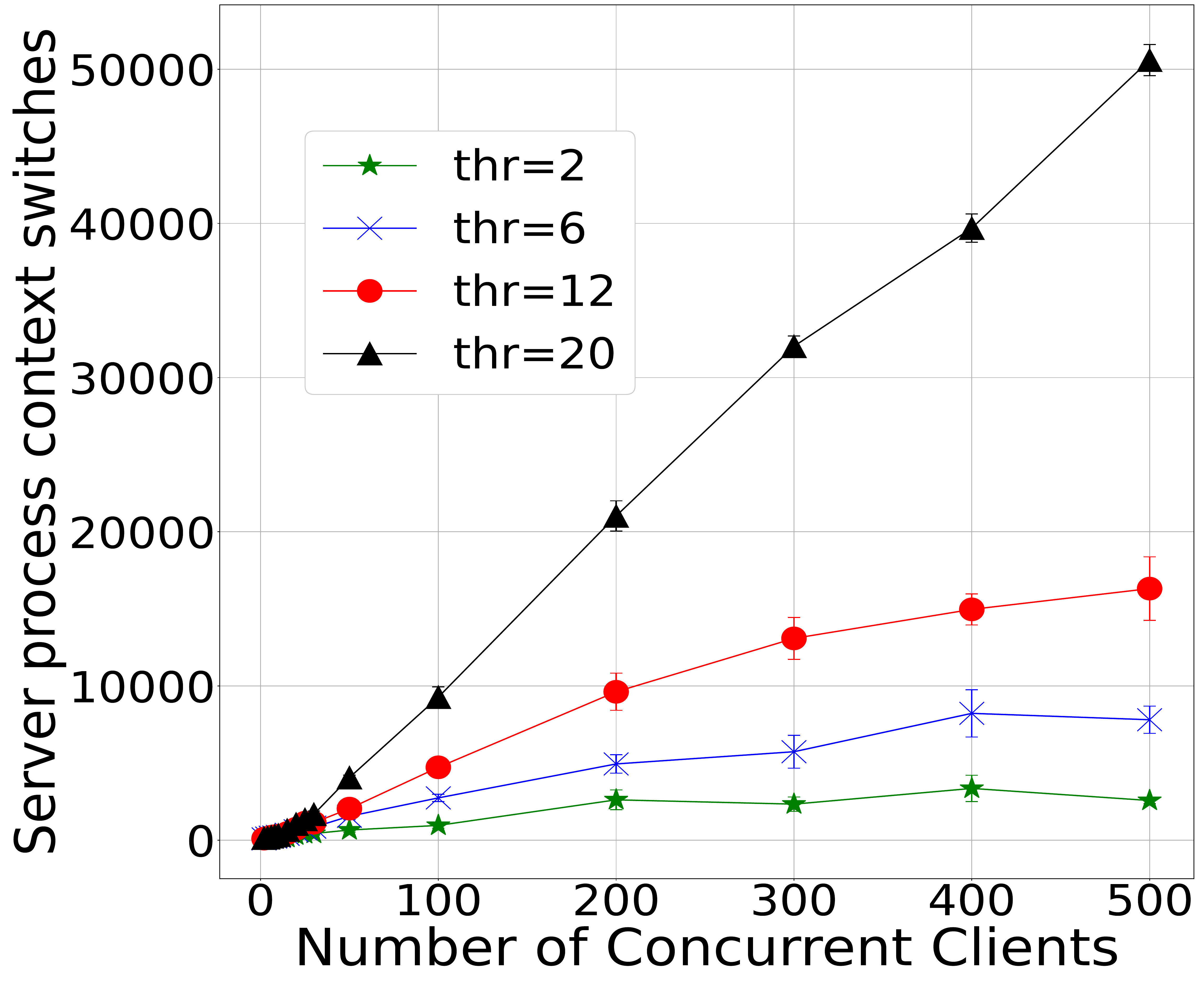}
        \caption[]%
        {{\small Server-side context switching}}    
        \label{fig:thr_ctxt}
    \end{subfigure}
    \vskip\baselineskip
    \caption{Operational benchmarking of the XRF server and client in a multi-threaded environment}
    \label{fig:thrall}
\end{figure}

We use end-to-end operational throughput on the server side and end-to-end latency on the client side for performance measurements. Our raw metrics are the system wall and CPU times. The former corresponds to the total operation time and the latter is the active processing time spent across the CPUs for all the threads. The results are given in Fig.~\ref{fig:thrall}.

In each experiment, we instantiate the given number of concurrent clients from separate containerized environments for end-to-end benchmarking of the preliminary flows given in Fig.~\ref{fig:xrfc_s_prelim}. For measuring the server throughput, an internal wall clock timer is started inside the application with the arrival of the first user and stopped when the final user is fully processed.

We can see in Fig.~\ref{fig:thr_throu} that increasing the number of threads running on the XRF server significantly impacts the number of users processed per second, indicating concurrency among the modules. The throughput saturates after a while regardless of the number of threads, which is a result of increased processor scheduling explained in the next paragraph. The saturation can be seen by comparing the marginal increase from 12 to 20 threads as opposed to the increase in 2 to 6 threads.

A lower-level explanation for this is provided by analyzing the context switching measurements in Fig.~\ref{fig:thr_ctxt}. To obtain these numbers, we lookup the process ID (PID) of the XRF server process on the L-worker in Section ~\ref{sec:expsetup} in the Linux file system. Later, we isolate the individual thread IDs (TIDs) belonging to this PID and record the total context switches performed across all the threads. This shows that there is a considerable increase going from 12 to 20 threads, which creates increased scheduling overhead in the processors. 

\begin{figure}[b]
    \centering
    %\vspace{-4mm}
    \begin{subfigure}[b]{0.475\columnwidth}
        \centering
        \includegraphics[width=\textwidth,trim={0cm 0cm 22.9cm, 0cm},clip]{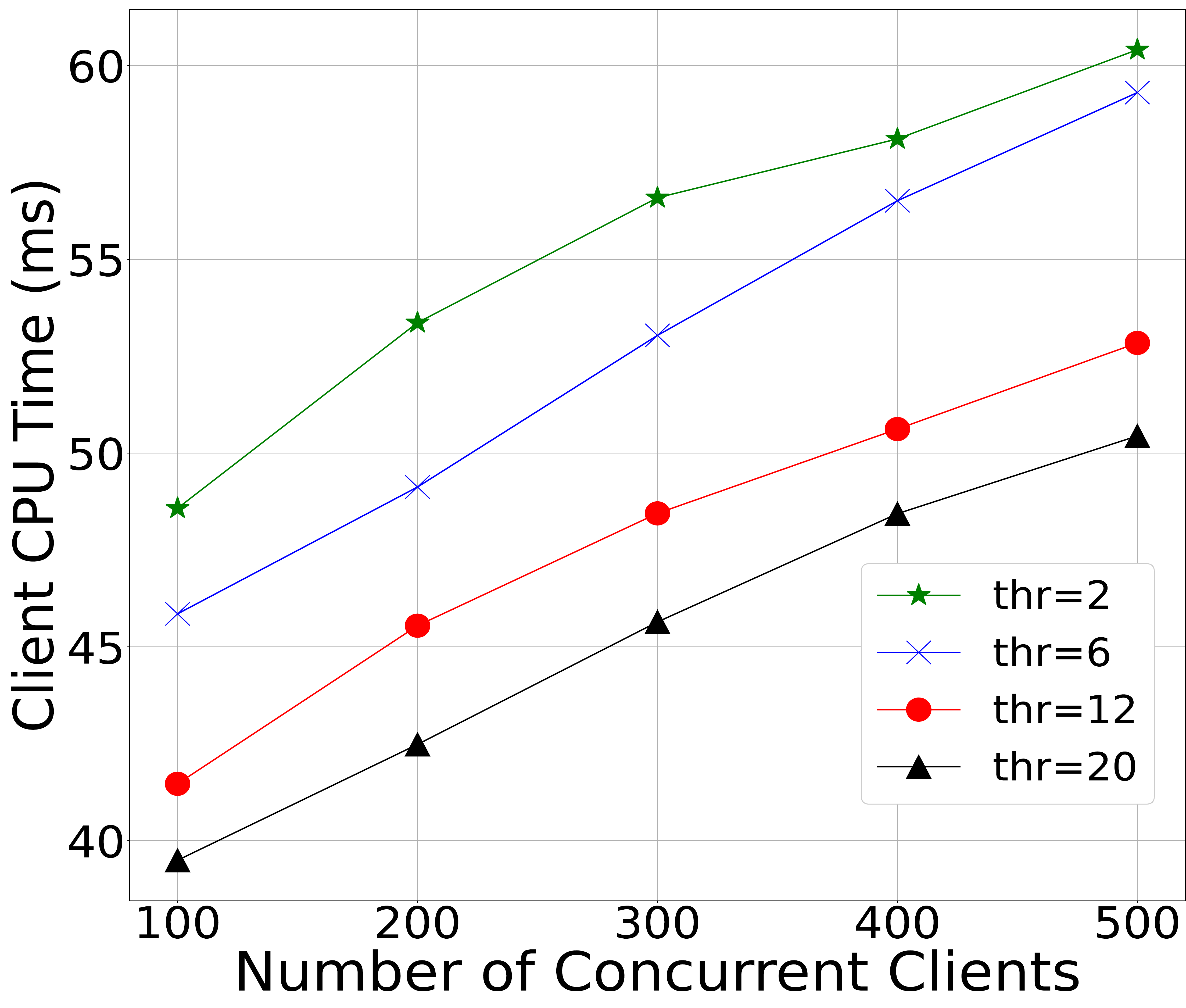}
        \caption[]%
        {{\small Total client CPU time}}    
        \label{fig:thr_cltct}
    \end{subfigure}
    \hfill
    \begin{subfigure}[b]{0.475\columnwidth}  
        \centering 
        \includegraphics[width=\textwidth,trim={0cm 0cm 22.8cm, 0cm},clip]{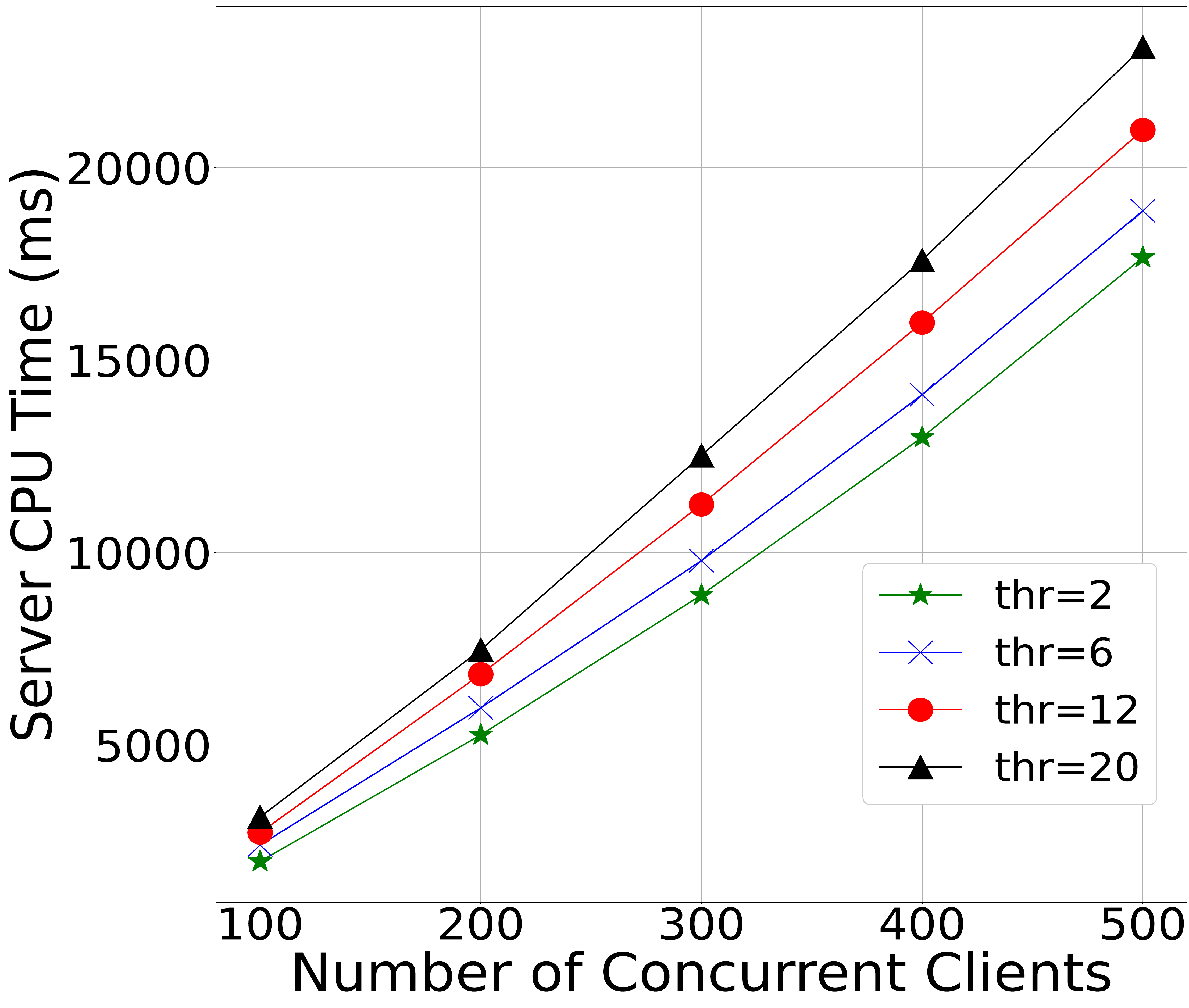}
        \caption[]%
        {{\small Total server CPU time}}    
        \label{fig:thr_serct}
    \end{subfigure}
    \vskip\baselineskip
    \caption{XRF server and client active CPU time measurements with multiple threads}
    \label{fig:thrcpu}
\end{figure}

The client end-to-end latency measurements given in Fig.~\ref{fig:thr_lat} complement the throughput results in Fig.~\ref{fig:thr_throu}. The latency is measured individually on each XRF client and averaged. With fewer threads on the XRF server, clients are suspended in scheduling, leading to higher wait times. 

In a multithreaded system, the combined CPU time will be higher than the overall wall time with multi-processing taking place over a set of CPUs. In Fig.~\ref{fig:thrcpu}, we show the total client and server CPU times with a varying number of threads.

The CPU time on the client side in Fig.~\ref{fig:thr_cltct} is inversely proportional to the number of threads on the XRF server. Fewer threads cause slower scheduling for the client, resulting in operational HTTP connections being kept open longer than necessary, leading to a slight overhead on the XRF clients. On the other hand, increasing the number of threads on the server allows for more processors to be engaged simultaneously which means the XRF server can handle tasks more efficiently, provided there are available CPUs.  

\begin{figure}[t]
    \centering
    \begin{subfigure}[b]{0.475\columnwidth}
        \centering
        \includegraphics[width=\textwidth,trim={0cm 0cm 21.6cm, 0cm},clip]{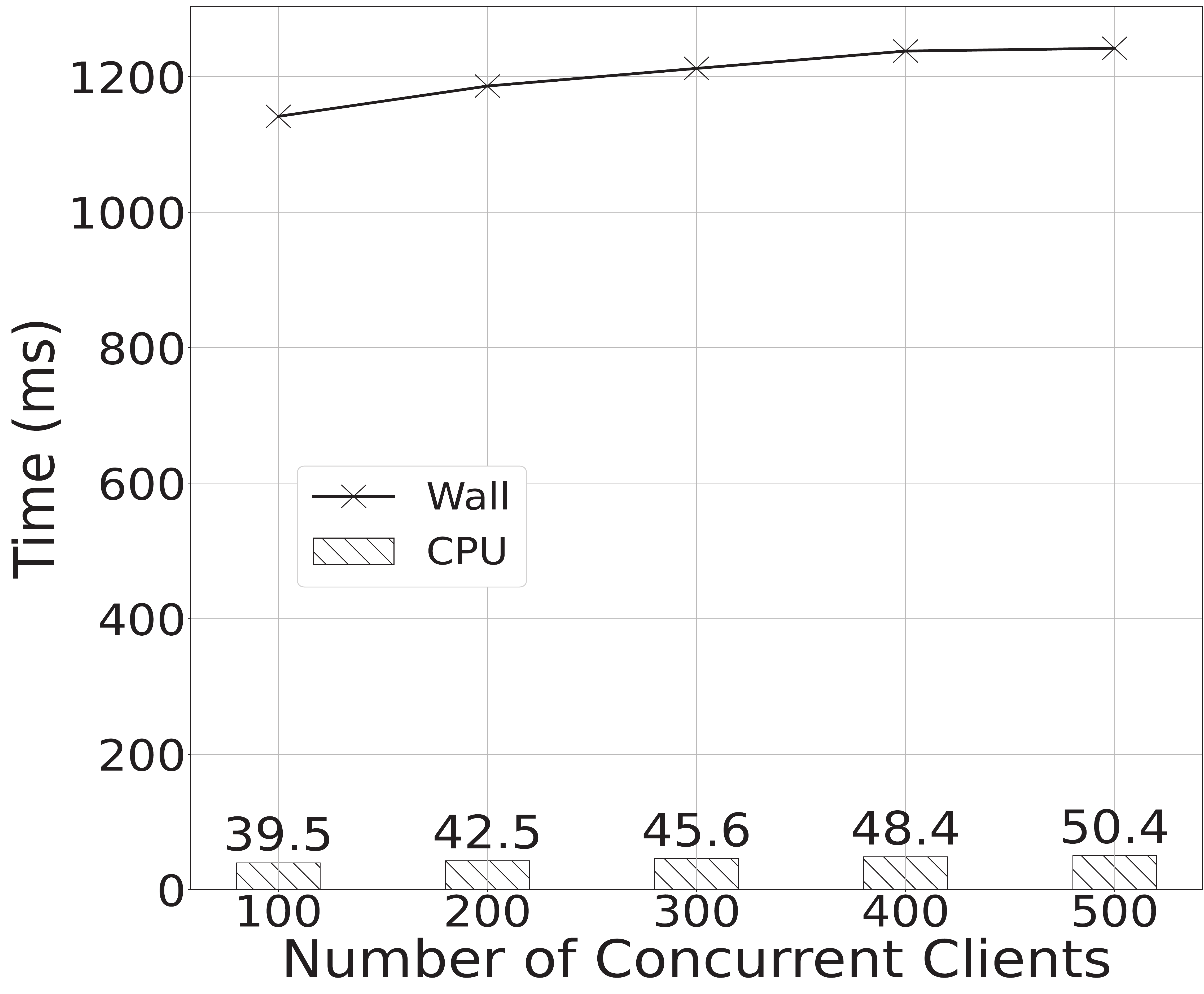}
        \caption[]%
        {{\small Client CPU vs wall time}}    
        \label{fig:cl_cpuwall}
    \end{subfigure}
    \hfill
    \begin{subfigure}[b]{0.475\columnwidth}  
        \centering 
        \includegraphics[width=\textwidth,trim={0cm 0cm 22cm, 0cm},clip]{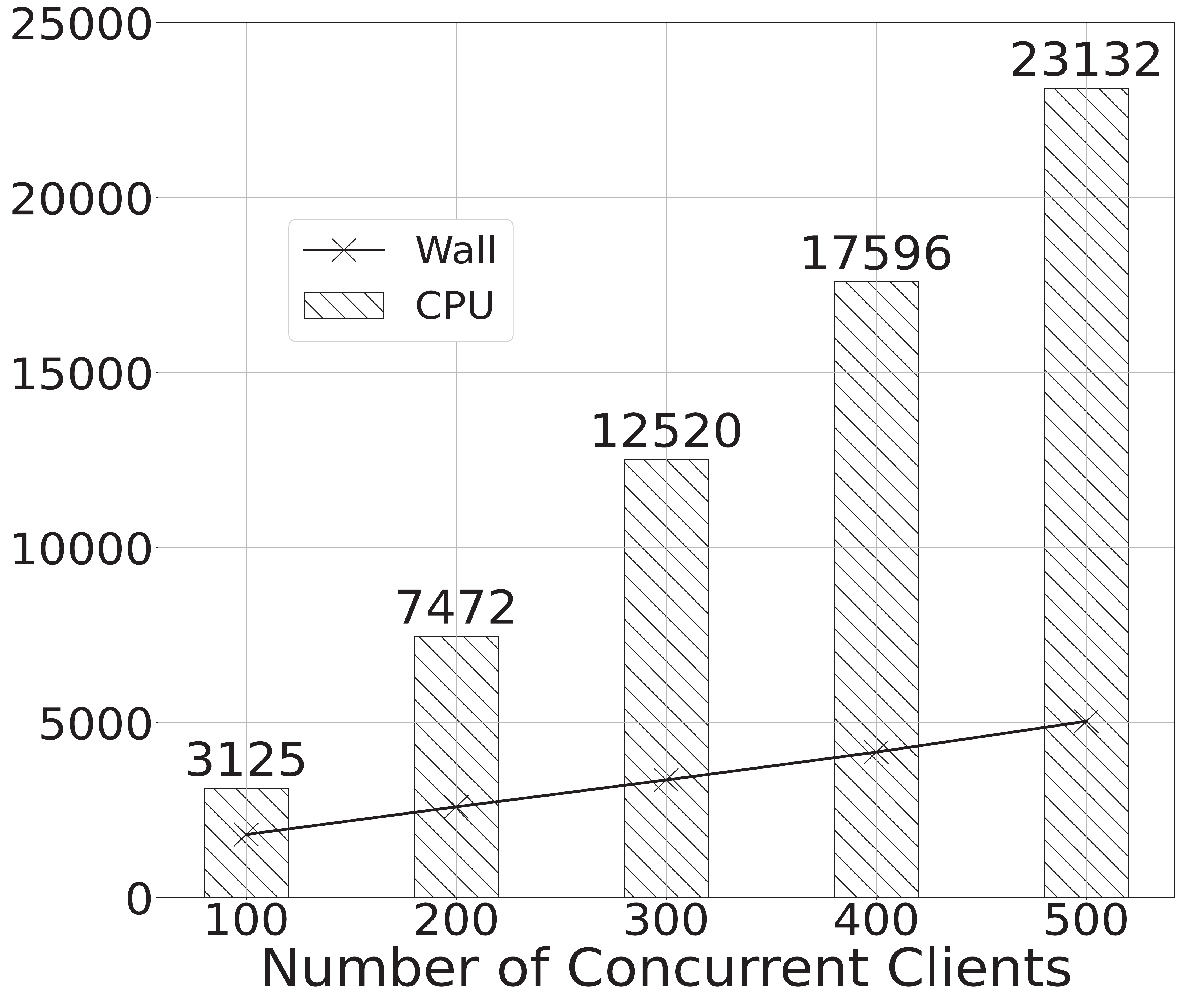}
        \caption[]%
        {{\small Server CPU vs wall time}}    
        \label{fig:serv_cpuwall}
    \end{subfigure}
    \vskip\baselineskip
    \caption{Comparing CPU and wall time on the XRF client and server for an HTTP server with 20 threads}
    \label{fig:serv_cl_wallcpu}
\end{figure}

To understand the difference between wall and CPU time, we compare the two for the XRF client and server in Fig.~\ref{fig:serv_cl_wallcpu} for a server with 20 threads. On the client side shown in Fig.~\ref{fig:cl_cpuwall}, CPU time is much lower than wall time because the majority of the latency is caused by processing on the server. This shows that the XRF client is a lightweight entity with minimal overhead, enabling xApps to efficiently scale in a production-grade framework. On the server side, however, as shown in Fig.~\ref{fig:serv_cpuwall}, CPU time is much higher than wall time because multiple processors are engaged simultaneously to execute different handlers within the XRF server. 

\begin{figure}[t]
    \centering
    \begin{subfigure}[b]{0.475\columnwidth}
        \centering
        \includegraphics[width=\textwidth,trim={0cm 0cm 22cm, 0cm},clip]{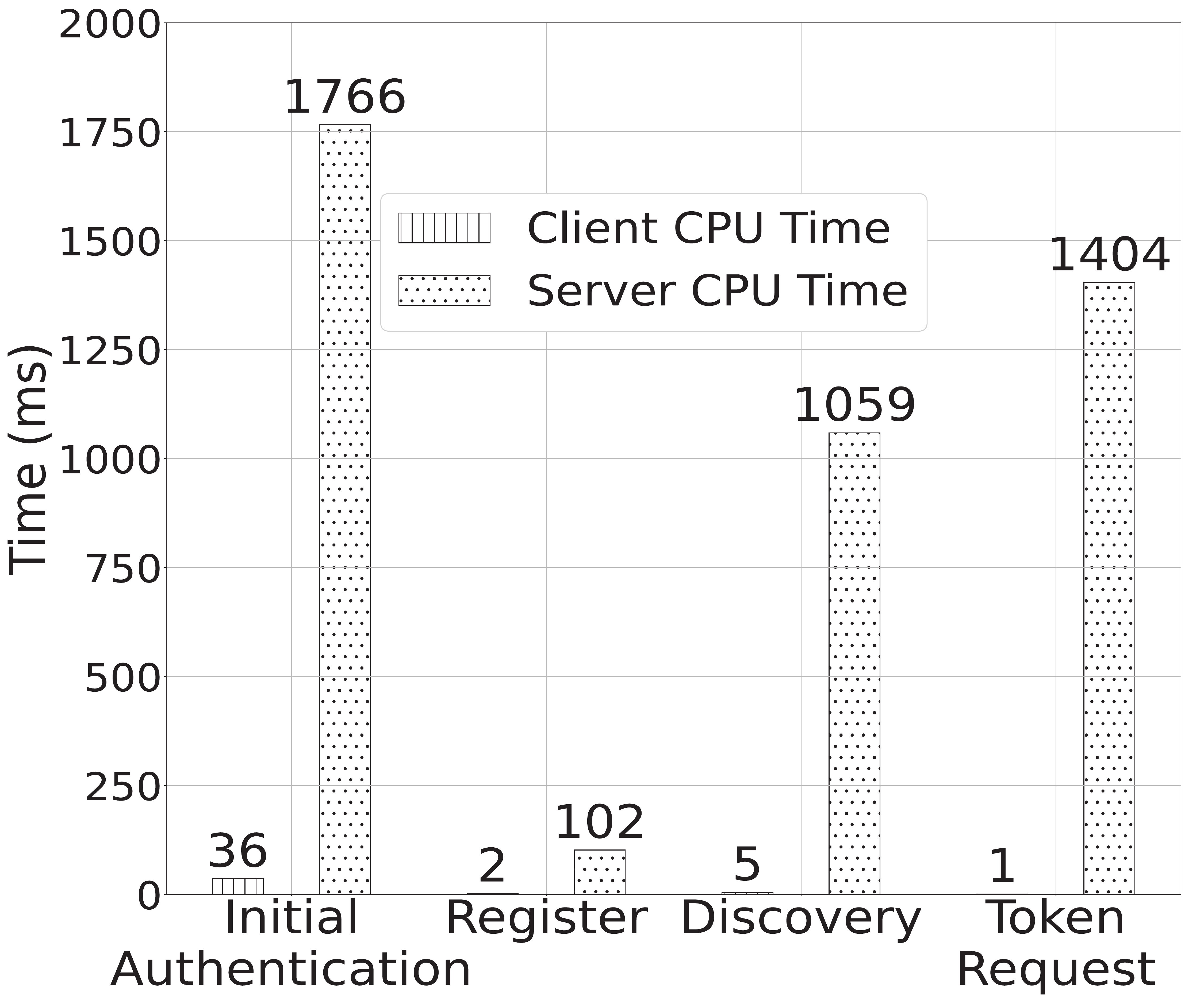}
        \caption[]%
        {{\small Results for flows in Fig.~\ref{fig:xrfc_s_prelim}}}    
        \label{fig:operations}
    \end{subfigure}
    \hfill
    \begin{subfigure}[b]{0.475\columnwidth}  
        \centering 
        \includegraphics[width=\textwidth,trim={0cm 0cm 22cm, 0cm},clip]{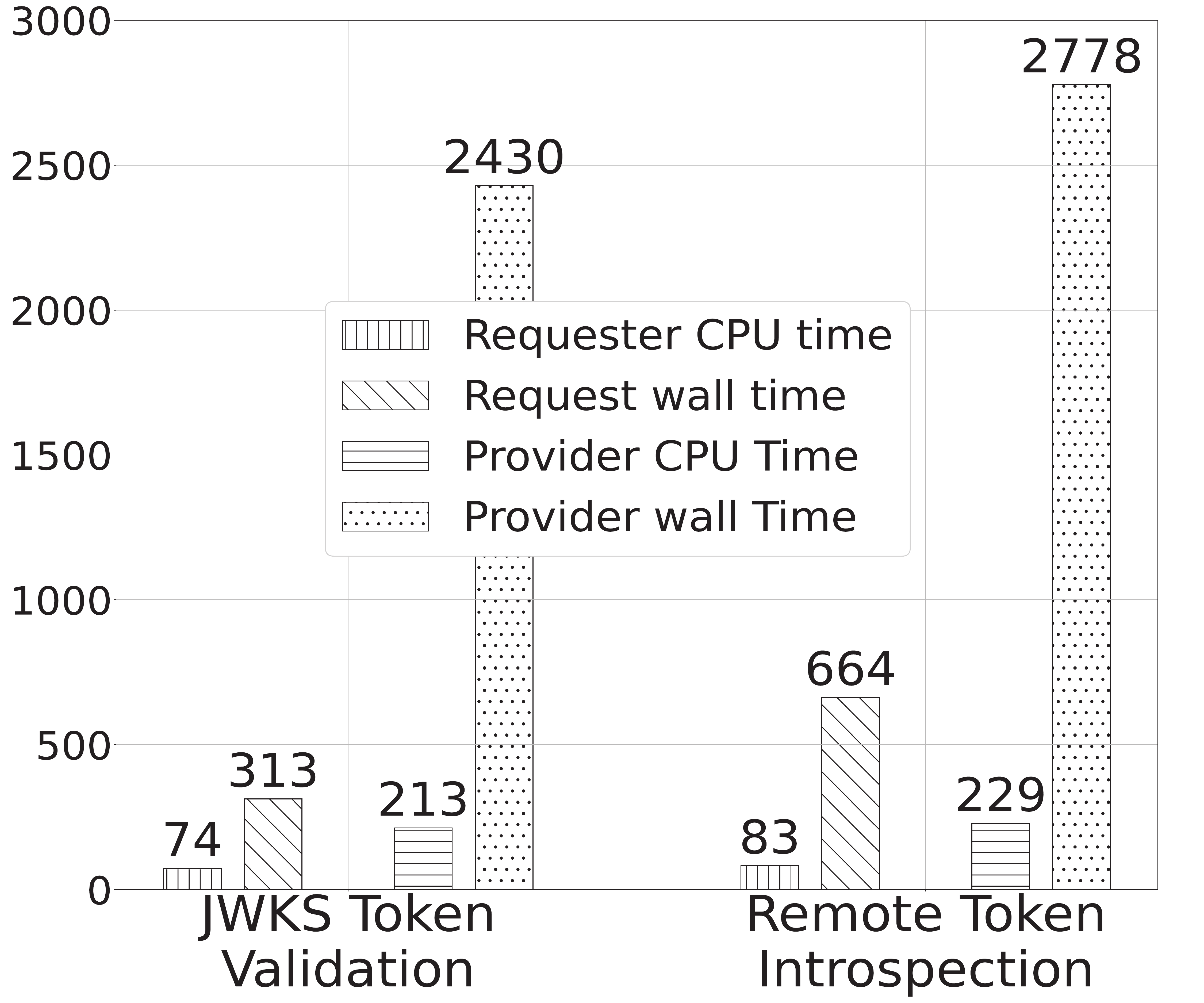}
        \caption[]%
        {{\small Results for flows in Fig.~\ref{fig:xrfc_s_cons}}}    
        \label{fig:operationstoken}
    \end{subfigure}
    \vskip\baselineskip
    \caption{Benchmarks for individual modules of XRF server and client operations}
    \label{fig:operationsall}
\end{figure}

Furthermore, we carried out operational benchmarks for the individual modules of the XRF framework in Fig.~\ref{fig:operationsall} with 50 concurrent clients. The analysis given in Fig.~\ref{fig:operations} illustrates the results for the initial authentication, registration, discovery, and access token modules. We can see that initial authentication, due to the cryptography operations, consumes the most resources both on the client and server sides. For the remainder of the operations, client side CPU time is negligible because the XRF client will only supply the XRF server with the desired HTTP request. On the server side, we can see that registration uses very little time, while discovery and access token requests are more considerable. The discovery operation needs to traverse a key-value map while the access token creation module performs the creation of a JWT payload as well as permission validation on the request.

In Fig.~\ref{fig:operationstoken}, we carry out a service request scenario with one XRF server and two XRF clients deployed where one of the clients is a service requester and the other a provider. In this experiment, the same requester sends 100 consecutive service requests to the provider with the same access token. We show the total CPU and wall time for the two token verification flows shown in Fig.~\ref{fig:xrfc_s_cons} for the 100 requests. 

Due to the overhead of contacting the XRF server each time, the wall time on the requester side is approximately twice as long with remote token introspection compared to JWKS token validation. On the provider side, we can see comparable wall times since both cases use the same verification method, regardless of whether the client or the server carries it out. 

\begin{comment}
We can summarize our results as follows.
\begin{itemize}
    \item End-to-end \textbf{server throughput}, \textbf{client latency} and \textbf{server context switches} across multiple threads.
    \item Total client and server \textbf{CPU times} with multithreading.
    \item \textbf{Comparison of CPU and wall time} for server and client.
    \item Benchmarks of \textbf{individual modules} in the framework.
\end{itemize}
\end{comment}

%% file: conc/texfile.tex
The OpenRAN framework is a next-generation mobile network platform aiming to achieve softwarization of the RAN and standardization of the interfaces across vendors. As with all new architectures, it expands the attack surface, requiring new security frameworks to address emerging threats. In this paper, we presented the system design, implementation, and evaluation of the XRF framework, which ensures the scalable authentication, authorization, and discovery of xApps within O-RAN, a software community realization of the OpenRAN initiative. XRF is built using proven and robust concepts tested in mobile app ecosystems, compatible with the containerized microservice model, offering seamless operational security for xApps. To fulfill the non-functional security requirements, we wrapped the deployment with the Linkerd service mesh to provide mTLS between entities. 
We conducted large-scale deployments in a HA Kubernetes cluster to evaluate XRF using operational benchmarks. Our results show that the XRF client has lightweight processing requirements, making it ideal for xApp adjacent deployment in a decentralized microservice environment. At the same time, the server-side handles the majority of the workload. Additionally, we demonstrate that the XRF server can scale in a multi-threaded environment while supporting concurrent clients with minimal overhead, trading off time for computing resources. 